\documentclass[prd,preprint,showpacs,preprintnumbers,nofootinbib,eqsecnum,superscriptaddress]{revtex4-1}

 \usepackage[dvips,final]{graphicx}
  \usepackage{amssymb}
   \usepackage{amsmath}
    \usepackage{amsfonts}
     \usepackage{epsfig}
      \usepackage{bm}

\usepackage{mathpazo}

\usepackage[section]{placeins}

\usepackage{multirow}
\usepackage{ctable}
\usepackage{booktabs}
\usepackage{array}
\usepackage{tabularx}
\usepackage{xcolor}
\usepackage{pstricks}

\newcommand{\dif}{\mathrm{d}}
\newcommand{\diff}[1]{\frac{\mathrm{d}#1}{#1}}

\begin{document}

\title{Single-diffractive production of charmed mesons at the LHC\\
within the \bm{$k_t$}-factorization approach}

\author{Marta {\L}uszczak}
\email{luszczak@univ.rzeszow.pl} 
\affiliation{University of Rzesz\'ow, PL-35-959 Rzesz\'ow, Poland}

\author{Rafa{\l} Maciu{\l}a}
\email{rafal.maciula@ifj.edu.pl} 
\affiliation{Institute of Nuclear Physics PAN, PL-31-342 Cracow, Poland}


\author{Antoni Szczurek\footnote{also at University of Rzesz\'ow, PL-35-959 Rzesz\'ow, Poland}}
\email{antoni.szczurek@ifj.edu.pl}
\affiliation{Institute of Nuclear Physics PAN, PL-31-342 Cracow, Poland}

\author{Maciej Trzebi{\'n}ski}
\email{maciej.trzebinski@ifj.edu.pl} 
\affiliation{Institute of Nuclear Physics PAN, PL-31-342 Cracow, Poland}

\date{\today}

\begin{abstract}
We discuss the single diffractive production of $c \bar c$ pairs and charmed mesons at the LHC. For a first time we propose a $k_t$-factorization approach to the diffractive processes. The transverse momentum dependent diffractive parton distributions are obtained from standard (collinear) diffractive parton distributions used in the literature. In this calculation the transverse momentum of the pomeron is neglected  with respect to transverse momentum of partons entering the hard process. We also perform the first evaluation of the cross sections at the LHC using the diffractive transverse momentum dependent parton distributions. The results of the new approach are compared with those of the standard collinear one. Significantly larger cross sections are obtained in the $k_t$-factorization approach where some part of higher-order effects is effectively included. The differences between corresponding differential distributions are discussed. Finally, we present a feasibility study of the process at the LHC using proton tagging technique. The analysis suggests that the measurement of single diffractive charm production is possible using ATLAS and CMS/TOTEM detectors.
\end{abstract}


\maketitle

\section{Introduction}
Diffractive hadronic processes are a special class of production mechanisms when forward emitted proton(s) is (are) accompanied by a sizeable  (a few rapidity units) rapidity gap(s) starting from the most forward  (in rapidity) proton(s) towards midrapidities. If only one of the forward protons is required, such processes are called single-diffractive. These processes were studied theoretically in the so-called resolved pomeron model \cite{IS}. It was realized during the Tevatron studies that the model, previously used to describe deep-inelastic diffractive processes, must be corrected to take into account absorption effects related to hadron-hadron interactions. Such interactions, unavoidably present in hadronic collisions at high energies and not present in electron/positron induced processes, lead to a damping of the diffractive cross section defined above. In theoretical models this effect is taken into account by multiplying the diffractive cross section calculated using HERA diffractive PDFs by a phase-space independent factor called the gap survival probability -- $S_{G}$. Two theoretical groups specialize in calculating such probabilities  \cite{KMR2000, Maor2009} and provide numerical values of $S_{G}$ for various energies and different types of diffractive mechanisms. At high energies such factors, interpreted as probabilities, are very small (of the order of a few \%). This causes that the predictions of the diffractive cross sections are not as precise as those for the standard inclusive non-diffractive cases. This may become a challenge after a precise data from the LHC will become available.

Several processes have been considered theoretically and discussed in the literature. Studies were performed for single $W$ \cite{Collins,GolecBiernat:2011dz} and $Z$ \cite{CSS09} boson, di-jet \cite{Kramer}, direct photon \cite{Kohara:2015nda}, photon-jet \cite{Marquet:2013rja},  di-lepton \cite{Kopeliovich:1999ka,Kubasiak:2011xs}, $W^+ W^-$ \cite{Luszczak:2014mta}, $c \bar c$ \cite{LMS2011} and $b \bar b$ \cite{Goncalves:2015cik} pair production. This list in not complete and other processes are also possible.

From the experimental side there are also several interesting results published. For example: rapidity cross sections \cite{ATLAS_rapidity_cs}, production of diffractive minimum-bias \cite{ATLAS_minbias, CMS_minbias}, production of $W$ bosons \cite{Abe:1997jp, Abazov:2003ti} and di-jets \cite{Affolder:2000vb, Affolder:2001zn, ATLAS_diff_dijets, CMS_diff_dijets}.

In this paper we consider diffractive production of charm for which rather large cross section at the LHC are expected, even within the leading-order (LO) collinear approach \cite{Luszczak:2014cxa}. On the other hand, it was shown that for the inclusive non-diffractive charm production the LO collinear approach is rather poor approximation and higher-order corrections are crucial. In contrast, the $k_t$-factorization approach, which effectively includes  higher-order effects, gives a good description of the LHC data  for inclusive charm production at $\sqrt{s}$ = 7 TeV (see \textit{e.g}. Ref.~\cite{Maciula:2013wg}). This strongly suggests that application of the $k_t$-factorization approach to diffractive charm production would be useful. 

Besides, the dipole approach is also often used to calculate cross section  for diffractive processes. However, as we discussed in Ref.~\cite{Luszczak:2013cba}, it gives only a small fraction of the diffractive cross section for the charm production.

To sum up, the measurement of diffractive production of charm would be very useful to pin down the mechanism of diffractive production in general.  Since such a measurement seems important and useful, in the present paper we present also a feasibility study for the diffractive production of charm mesons within the ATLAS and CMS/TOTEM detectors.
 
\section{Formalism}

\begin{figure}[!htbp]
\begin{minipage}{0.49\textwidth}
 \centerline{\includegraphics[width=1.0\textwidth]{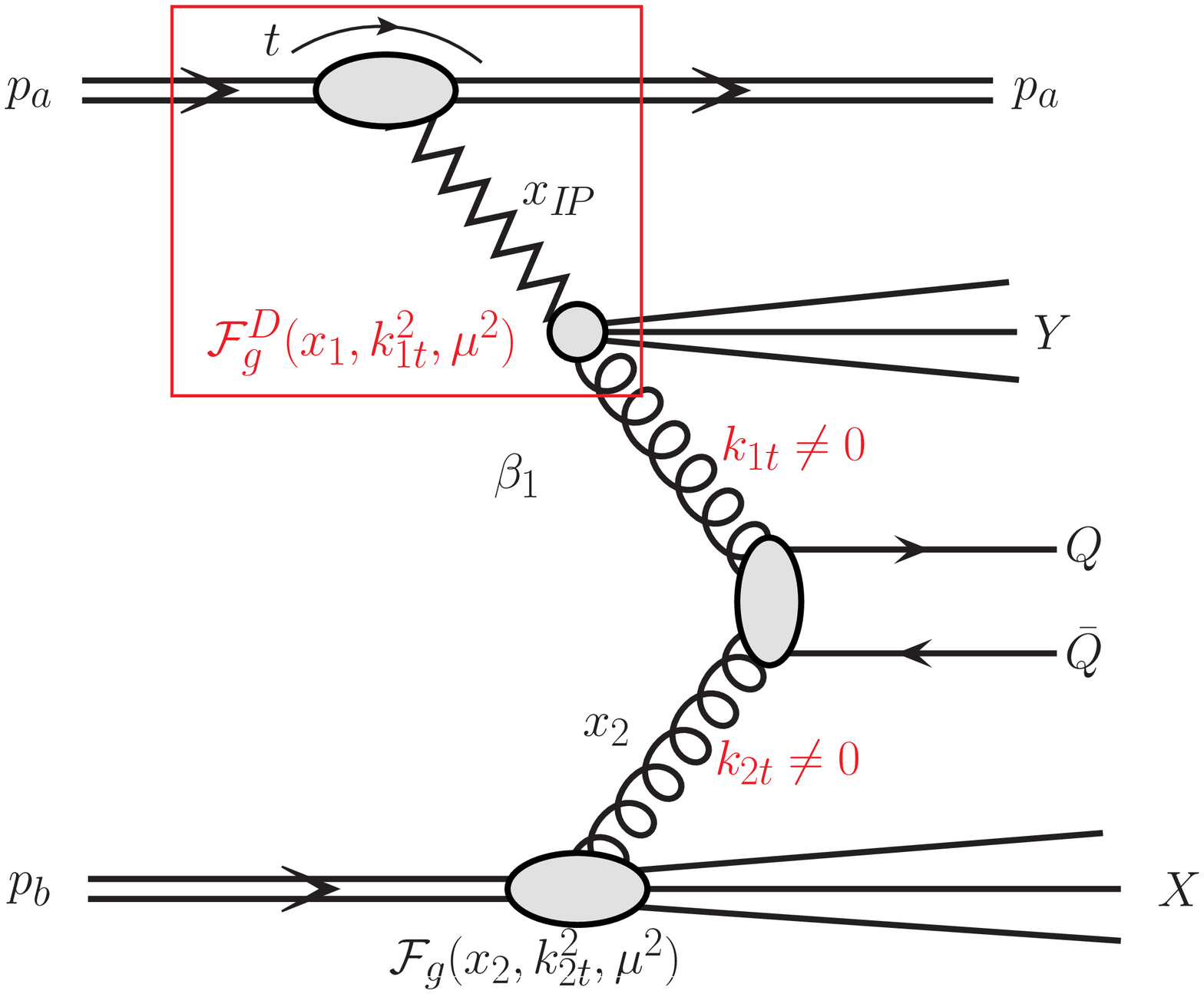}}
\end{minipage}
\begin{minipage}{0.49\textwidth}
 \centerline{\includegraphics[width=1.0\textwidth]{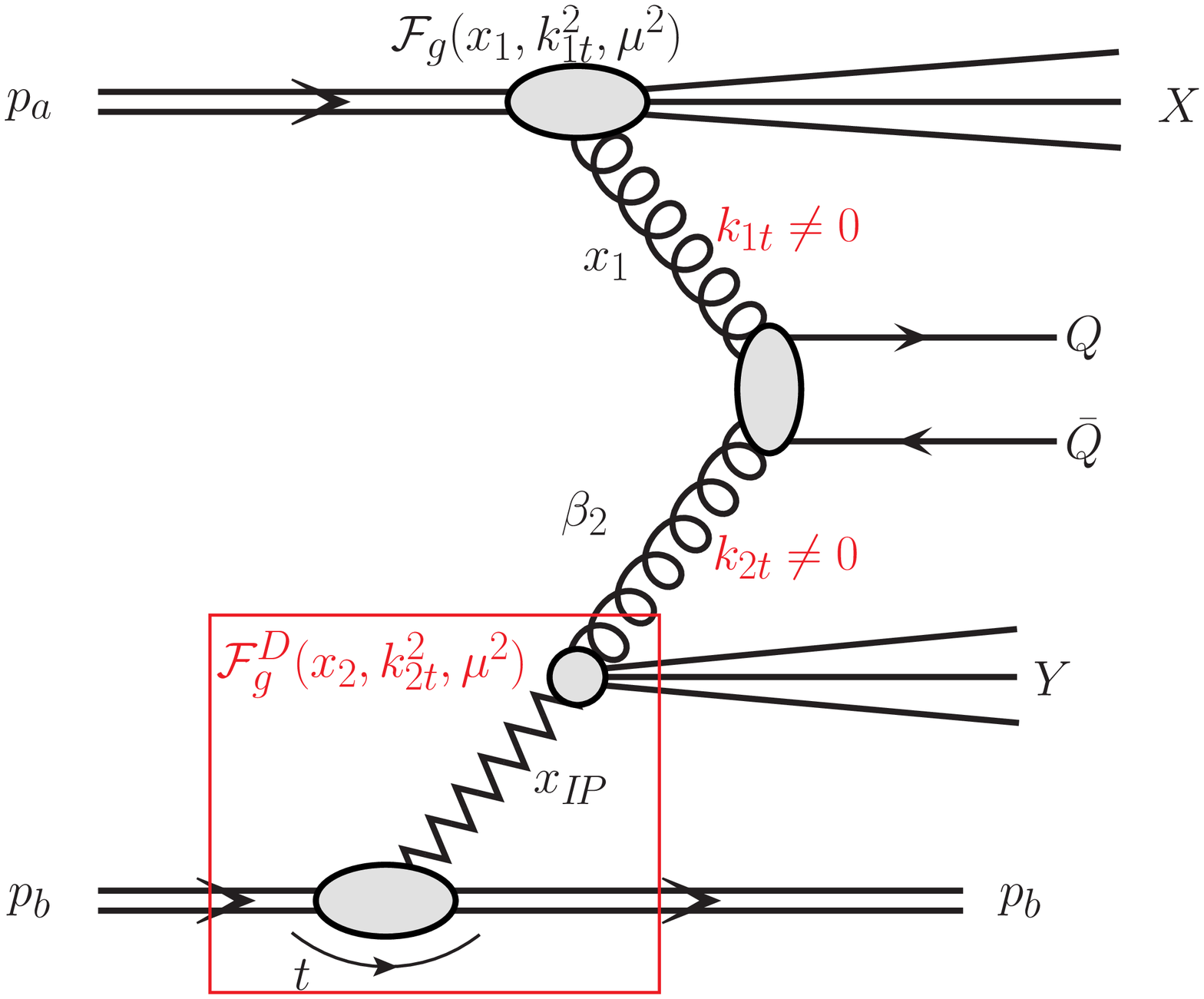}}
\end{minipage}
\caption{
\small A diagrammatic representation of the considered mechanisms of single-diffractive production of heavy quark pairs within the $k_{t}$-factorization approach.
}
 \label{fig:mechanism}
\end{figure}

A sketch of the underlying mechanism with the notation of kinematical variables or some theoretical ingredients used in the theoretical formalism is shown in Fig.~\ref{fig:mechanism}. Here, we propose extension of the standard resolved pomeron model \cite{IS} based on the LO collinear approach by adopting a framework of the $k_{t}$-factorization as an effective way to include higher-order corrections. Such an approach is very usefull \textit{e.g.} in the studies of kinematical correlations \cite{Maciula:2013wg}. According to this model the cross section for a single-diffractive production of charm quark-antiquark pair, for both considered diagrams (left and right panel of Fig.~\ref{fig:mechanism}), can be written as:
\begin{eqnarray}
d \sigma^{SD(a)}({p_{a} p_{b} \to p_{a} c \bar c \; X Y}) &=&
\int d x_1 \frac{d^2 k_{1t}}{\pi} d x_2 \frac{d^2 k_{2t}}{\pi} \; d {\hat \sigma}({g^{*}g^{*} \to c \bar c }) \nonumber \\
&& \times \; {\cal F}_{g}^{D}(x_1,k_{1t}^2,\mu^2) \cdot {\cal F}_{g}(x_2,k_{2t}^2,\mu^2) ,
\label{SDA_formula}
\end{eqnarray}
\begin{eqnarray}
d \sigma^{SD(b)}({p_{a} p_{b} \to c \bar c p_{b} \; X Y}) &=&
\int d x_1 \frac{d^2 k_{1t}}{\pi} d x_2 \frac{d^2 k_{2t}}{\pi} \; d {\hat \sigma}({g^{*}g^{*} \to c \bar c }) \nonumber \\
&& \times \; {\cal F}_{g}(x_1,k_{1t}^2,\mu^2) \cdot {\cal F}_{g}^{D}(x_2,k_{2t}^2,\mu^2),
\label{SDB_formula}
\end{eqnarray}
where ${\cal F}_{g}(x,k_{t}^2,\mu^2)$ are the "conventional" unintegrated ($k_{t}$-dependent) gluon distributions (UGDFs) in the proton and ${\cal F}_{g}^{D}(x,k_{t}^2,\mu^2)$ are their diffractive counterparts -- which we will call here diffractive UGDFs (dUGDFs). The latter can be interpreted as a probability of finding a gluon with longitudinal momentum fraction $x$ and transverse momentum (virtuality) $k_{t}$ at the factorization scale $\mu^{2}$ assuming that the proton which lost a momentum fraction $x_{I\!P}$ remains intact.

In the approach applied here, we neglect a possible influence of the pomeron/reggeon transverse momentum on the transverse momentum of the initial off-shell gluon from pomeron/reggeon. The effect is assumed to be negligible since the transverse momenta of incident gluon are typically larger (or much larger) than the transverse momenta of pomeron/reggeon (equal to the transverse momenta of outgoing proton). We will come back to the issue when presenting numerical results. 

The partonic cross section for the considered hard scattering reads:
\begin{eqnarray}
d {\hat \sigma}({g^{*}g^{*} \to c \bar c }) &=& \frac{d^3 p_1}{2 E_1 (2 \pi)^3} \frac{d^3 p_2}{2 E_2 (2 \pi)^3}
(2 \pi)^2 \delta^{2}(p_1 + p_2 - k_1 - k_2) \times \overline{|{\cal M}_{g^* g^* \to c \bar c}(k_{1},k_{2})|^2} \;, \nonumber \\
\label{elementary_cs}
\end{eqnarray}
where $p_1, E_1$ and $p_2, E_2$ are the momenta and energies of outgoing $c$ and $\bar c$, respectively, and $\overline{|{\cal M}_{g^* g^* \to c \bar c}(k_{1},k_{2})|^2}$ is the off-shell matrix element for the $g^* g^* \to c \bar c$  sub-process.

According to the so-called proton-vertex factorization, the diffractive collinear gluon distribution in the proton can be written in the form where the variables related with the proton kinematics are separated from those connected with the hard interaction:
\begin{eqnarray}
g^D(x,\mu^2) = \int d x_{I\!P} d\beta \, \delta(x-x_{I\!P} \beta) 
g_{I\!P} (\beta,\mu^2) \, f_{I\!P}(x_{I\!P}) \, 
= \int_x^{x^{max}} {d x_{I\!P} \over x_{I\!P}} \, f_{I\!P}(x_{I\!P})  
g_{I\!P}({x \over x_{I\!P}}, \mu^2) , \nonumber \\  
\end{eqnarray} 
where $\beta = \frac{x}{x_{I\!P}}$ is the longitudinal momentum fraction of pomeron carried by gluon and the flux of pomerons may be taken as:
\begin{equation}
f_{I\!P}(x_{I\!P}) = \int_{t_{min}}^{t_{max}} dt \, f(x_{I\!P},t).
\end{equation}

The unintegrated ($k_t$-dependent) diffractive gluon distributions in the proton can be easily obtained from the collinear diffractive PDFs by applying the Kimber-Martin-Ryskin (KMR) approach \cite{KMR}. According to this procedure, the diffractive unintegrated gluon distribution is given by the following formula:
\begin{eqnarray} \label{eq:UPDF}
  f^{D}_{g}(x,k_t^2,\mu^2) &\equiv& \frac{\partial}{\partial \log k_t^2}\left[\,g^{D}(x,k_t^2)\,T_g(k_t^2,\mu^2)\,\right] =
  T_g(k_t^2,\mu^2)\,\frac{\alpha_S(k_t^2)}{2\pi}\, \times  \\  
&& \!\!\!\int_x^1\! d z \left[\sum_q P_{gq}(z)\frac{x}{z}\;q^{D}\!\left(\frac{x}{z},k_t^2\right) + P_{gg}(z)\frac{x}{z}\;g^{D}\!\left(\frac{x}{z},k_t^2\right)\Theta\left(\Delta - z\right)\right], \nonumber
\end{eqnarray}
where $g^{D}$ and $q^{D}$ are the collinear diffractive PDFs in the proton and can be taken \textit{e.g.} from the H1 Collaboration analysis of diffractive dijets and diffractive structure function \cite{H1}. The $P_{gq}$ and $P_{gg}$ are the usual unregulated LO DGLAP splitting functions. The Heaviside step function $\Theta$ implies the angular-ordering constraint of the phase space $\Delta = \mu/( \mu + k_{t})$ for gluon emission particularly to the last evolution step to regulate the soft gluon singularities. The above definition is fully satisfied for $k_t > \mu_0$, where $\mu_0\sim 1$~GeV is the minimum scale for which DGLAP evolution of the collinear diffractive PDFs is valid. The Sudakov form factor $T_g(k_t^2,\mu^2)$ is responsible for virtual loop corrections and gives the probability of evolving from a scale $k_t$ to a scale $\mu$ without any new parton emissions. In the simplified form it may be written as:
\begin{equation}
  T_g(k_t^2,\mu^2) = \exp\left(-\int_{k_t^2}^{\mu^2}\!\diff{\kappa_t^2}\,\frac{\alpha_S(\kappa_t^2)}{2\pi}\,\left( \int_{0}^{1-\Delta}\!\dif{z}\;z \,P_{gg}(z) + n_F\,\int_0^1\!\dif{z}\,P_{qg}(z)\right)\right),
\end{equation}
where $n_F$ is the number of quark-antiquark flavours into which the gluon may split. Due to the presence of the Sudakov form factor in the KMR prescription only last emission generates transverse momentum of the incoming gluons. The unique feature of the KMR model of UGDF is providing possibility for the emission of at most one additional gluon. Therefore, one can expect that the KMR model may include in an effective way higher-order corrections to heavy quark production cross section.

Because of the UGDF definition in the KMR approach one needs to also apply the following transformation:  
\begin{eqnarray}
{\cal F}^{D}_{g}(x,k_{t}^2,\mu^2) \equiv \frac{1}{k_t^2}\,f^{D}_{g}(x,k_t^2,\mu^2)
\; .
\end{eqnarray}
Then the normalisation condition for diffractive unintegrated gluon distribution:
\begin{equation} \label{eq:norm}
  g^{D}(x,\mu^2) = \int_0^{\mu^2}\! d k_t^2\,f^{D}_{g}(x,k_t^2,\mu^2),
\end{equation}
is exactly satisfied if one defines:
\begin{equation} \label{eq:smallkt}
  \left.{\cal F}^{D}_{g}(x,k_{t}^2,\mu^2)\right\rvert_{k_t<\mu_0} = \frac{1}{\mu_0^2}\,g^{D}(x,\mu_0^2)\,T_g(\mu_0^2,\mu^2),
\end{equation}
so that the density of gluons in proton is constant for $k_t<\mu_0$
at fixed $x$ and $\mu$.

In this paper, the conventional non-diffractive KMR UGDFs are calculated from the MSTW2008lo collinear PDFs \cite{Martin:2009iq}. In the perturbative part of calculations we use running LO coupling constant $\alpha_{S}(\mu_{R}^{2})$ as implemented in the MSTW2008 code, the charm quark mass of $m_{c} = 1.5$ GeV and the renormalization and factorization scales equal to the transverse mass of charm quarks/antiquarks $\mu^{2} = \mu_{R}^{2} = \mu_{F}^{2} = \frac{m_{1t}^{2} + m_{1t}^{2}}{2}$, where $m_{t} = \sqrt{p_{t}^{2} + m_{c}^{2}}$.

\section{First Numerical Results}

\begin{figure}[!htbp]
\begin{minipage}{0.47\textwidth}
 \centerline{\includegraphics[width=1.0\textwidth]{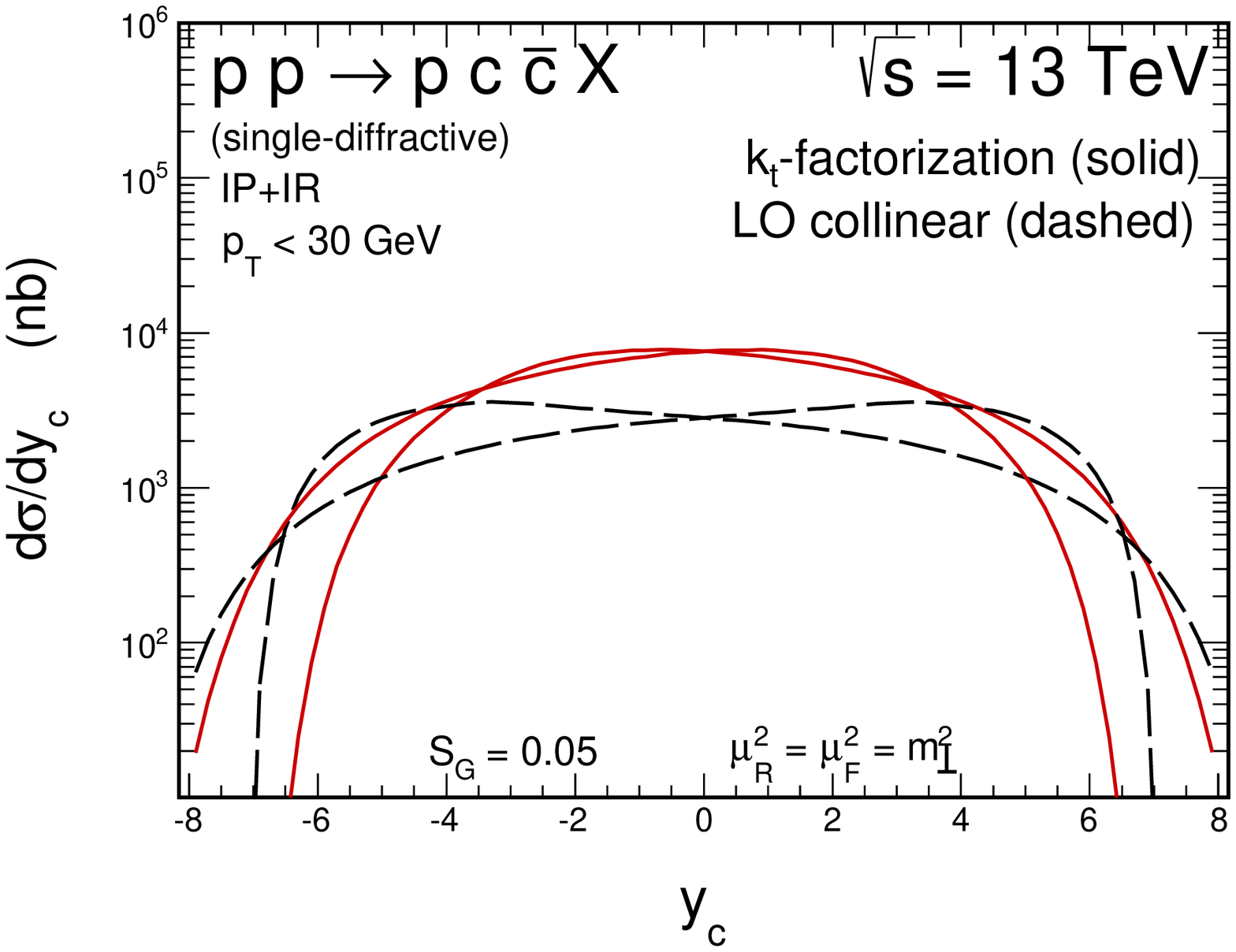}}
\end{minipage}
\hspace{0.5cm}
\begin{minipage}{0.47\textwidth}
 \centerline{\includegraphics[width=1.0\textwidth]{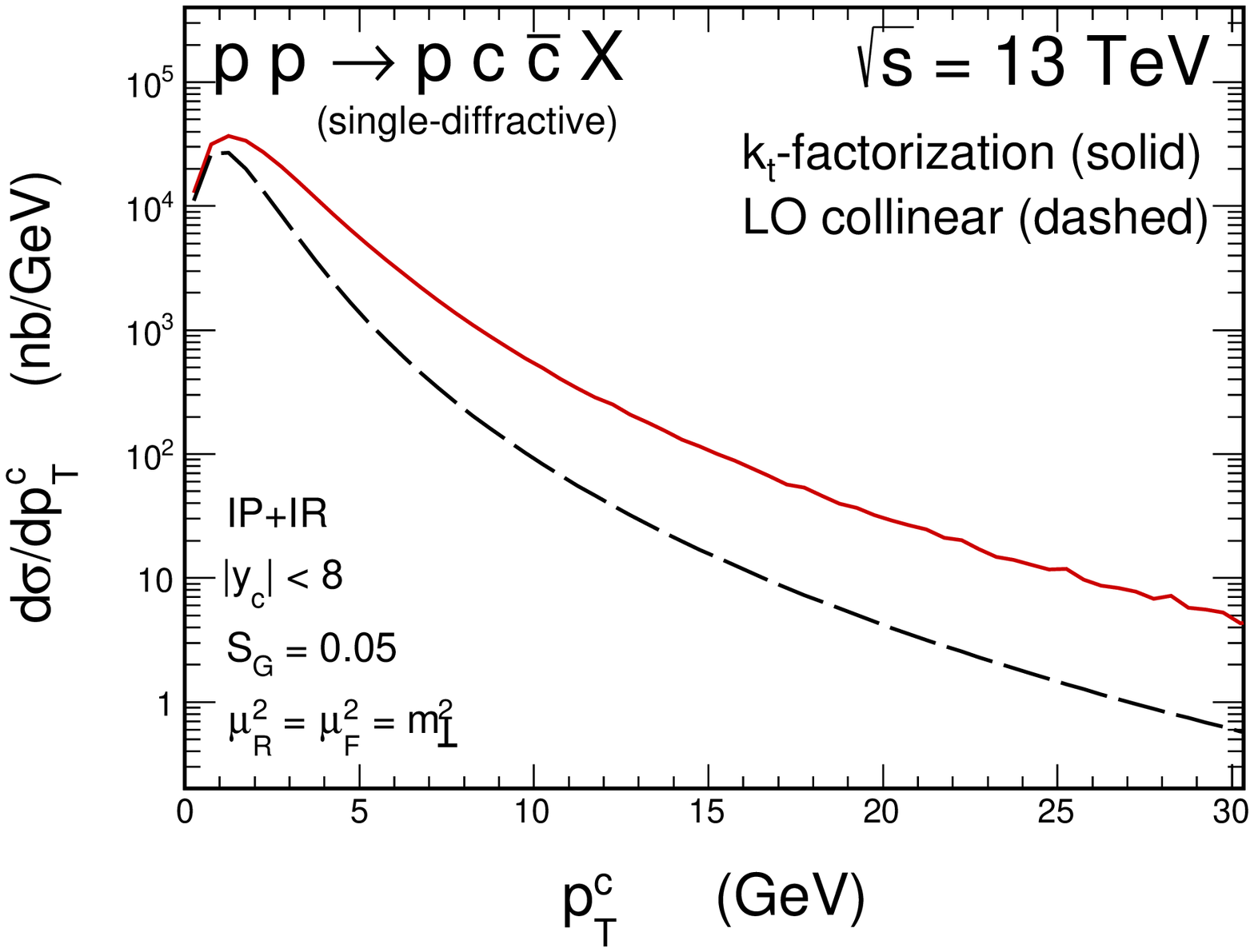}}
\end{minipage}
   \caption{
\small Rapidity (left panel) and transverse momentum (right panel) distributions of $c$ quarks (antiquarks) for a single-diffractive production at $\sqrt{s} = 13$ TeV. Components of the $g(I\!P)\operatorname{-}g(p)$, $g(p)\operatorname{-}g(I\!P)$, $g(I\!R)\operatorname{-}g(p)$, $g(p)\operatorname{-}g(I\!R)$ mechanisms are included. Results for the LO collinear (black long-dashed) and the $k_{t}$-factorization approach (red solid) are shown separately. Details are specified in the figure legends.
}
 \label{fig:ypt_kTcoll}
\end{figure}

In Fig.~\ref{fig:ypt_kTcoll} we show rapidity (left panel) and transverse momentum (right panel) distribution of $c$ quarks (antiquarks) for the single diffractive production at $\sqrt{s} = 13$ TeV in proton-proton scattering. The leading $g(I\!P)\operatorname{-}g(p)$, \textit{i.e.} gluon in pomeron - gluon in proton, (or $g(p)\operatorname{-}g(I\!P)$) and the subleading $g(I\!R)\operatorname{-}g(p)$, \textit{i.e.} gluon in reggeon - gluon in proton, (or $g(p)\operatorname{-}g(I\!R)$) components are included. We limit the range of the momentum fractions $x_{I\!P}$ and $x_{I\!R}$ in the numerical calculations to $x_{I\!P}, x_{I\!R} < 0.15$ which reflects the maximal experimental coverage at the LHC. The calculation assumes Regge factorization, which is known to be violated in the hadron-hadron collisions. This is due to the fact that soft interactions lead to an extra production of particles which fill the rapidity gaps related to pomeron/reggeon exchange. Therefore, in the calculations we use the gap survival probability ($S_G$ = 0.05) to effectively include these effects. Distributions calculated within the LO collinear factorization (black long-dashed lines) and for the $k_{t}$-factorization approach (red solid lines) are shown separately. We see significant differences between the both approaches, which are consistent with similar studies of standard non-diffractive charm production (see \textit{e.g}. Ref.~\cite{Maciula:2013wg}). Here we confirm that the higher-order corrections are very important also for the diffractive production of charm quarks. Predictions within the $k_{t}$-factorization give a significantly larger differential cross section in the whole $p_{t}$ and $y$ ranges, except for a very small transverse momenta and extremely forward/backward rapidities.
 
\begin{figure}[!htbp]
\begin{minipage}{0.47\textwidth}
 \centerline{\includegraphics[width=1.0\textwidth]{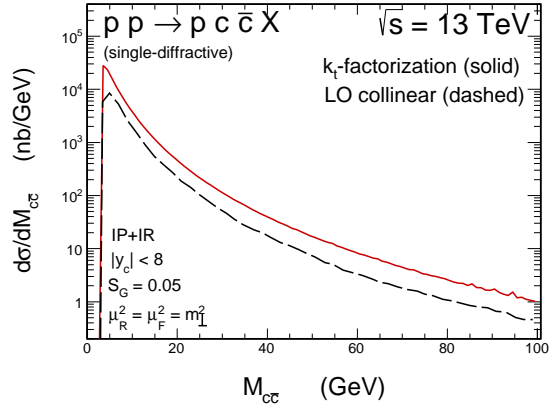}}
\end{minipage}
   \caption{
\small Invariant mass distributions of $c \bar c$ pairs for a single-diffractive production at $\sqrt{s} = 13$ TeV. Here, the charm quark rapidity and transverse momentum were limited to $|y| < 8$ and $0 < p_{t} < 30$~GeV, respectively. 
}
 \label{fig:Minv_kTcoll}
\end{figure}

In Fig.~\ref{fig:Minv_kTcoll} we show the invariant mass distribution of $c \bar c$ pairs for a single-diffractive production at $\sqrt{s} = 13$ TeV. The shapes of the distributions obtained for the LO collinear and the $k_{t}$-factorization approach are rather similar. However, the latter gives larger cross section approximately by a factor of almost 3.

\begin{figure}[!htbp]
\begin{minipage}{0.47\textwidth}
 \centerline{\includegraphics[width=1.0\textwidth]{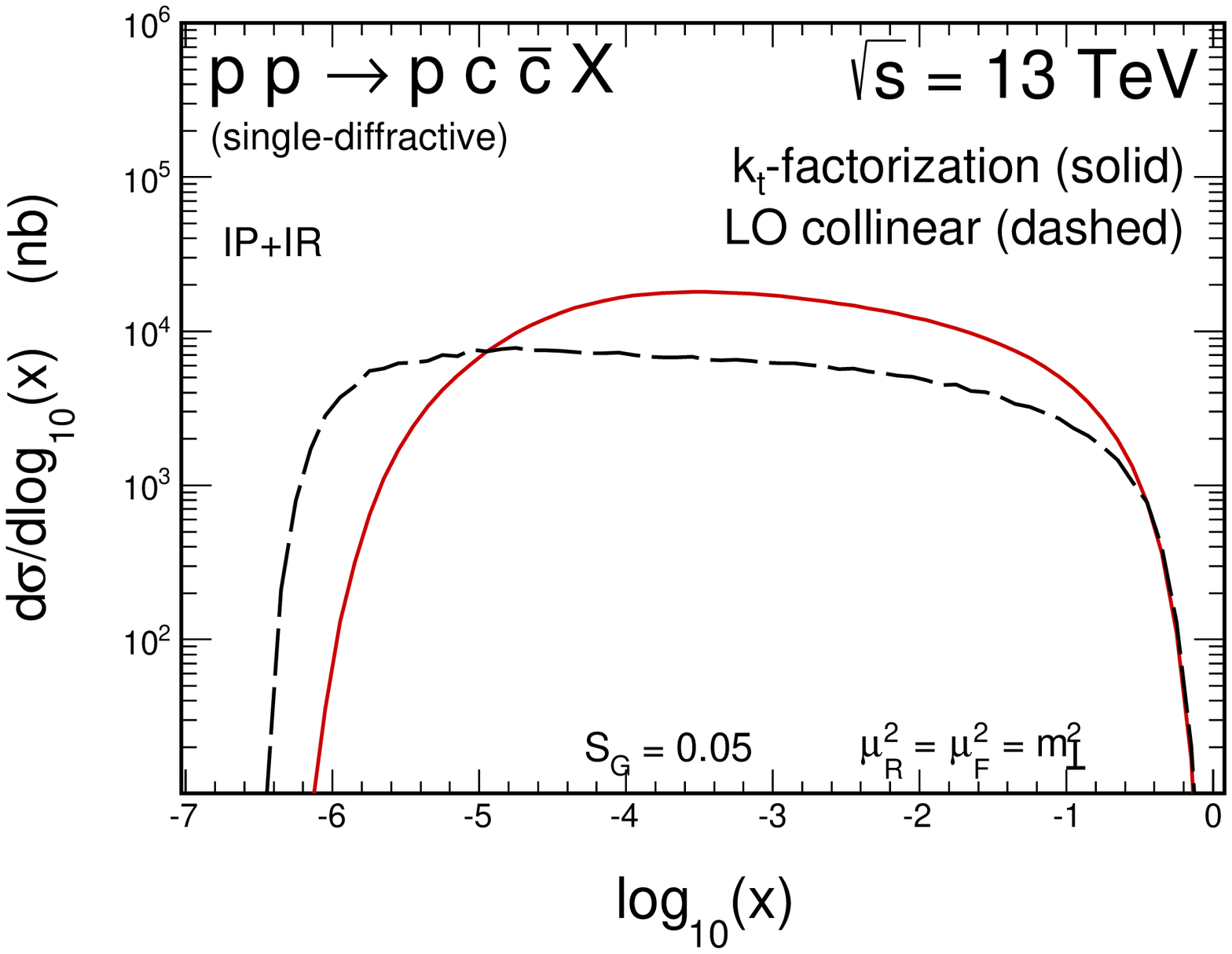}}
\end{minipage}
\hspace{0.5cm}
\begin{minipage}{0.47\textwidth}
 \centerline{\includegraphics[width=1.0\textwidth]{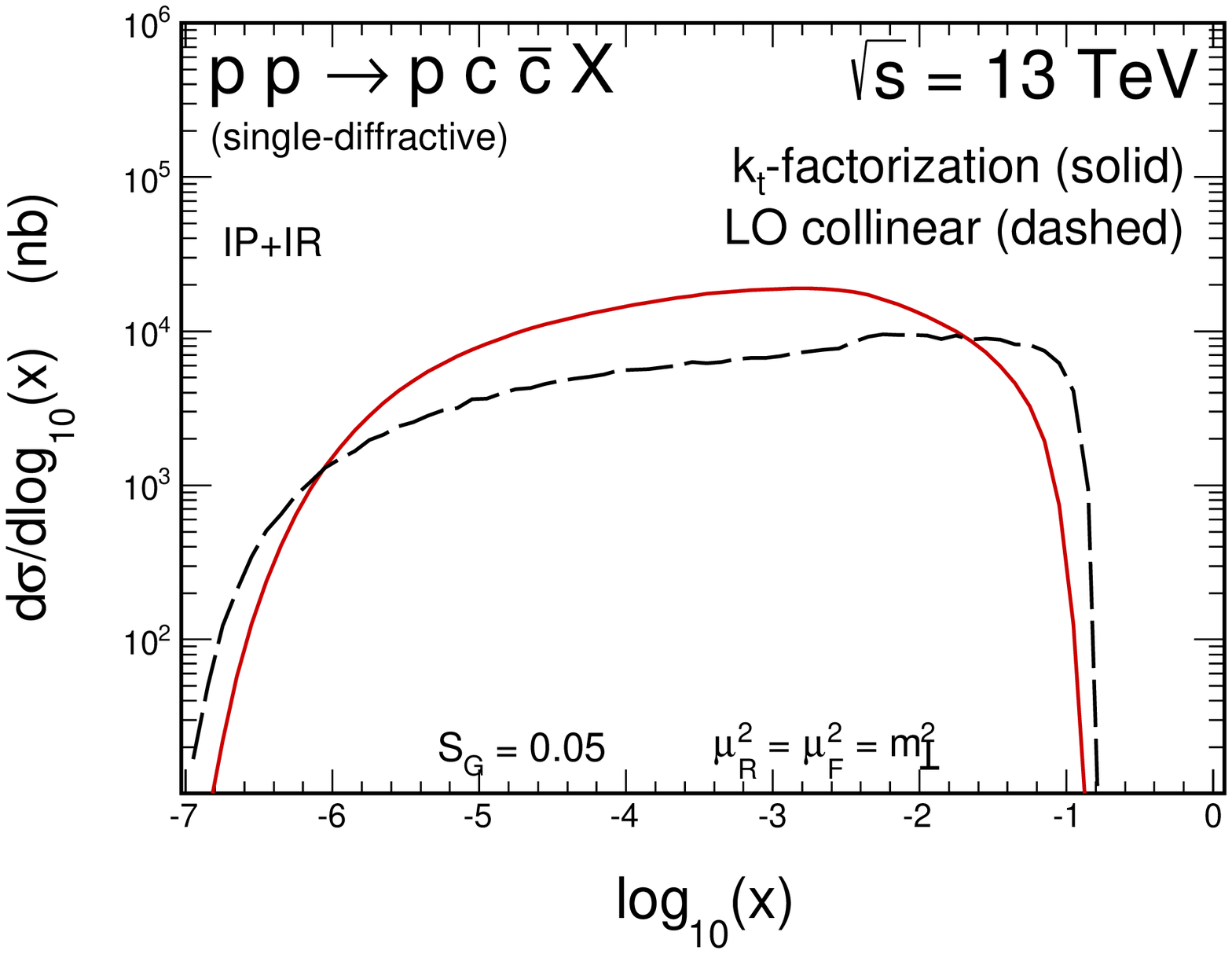}}
\end{minipage}
   \caption{
\small The differential cross section as a function of $\log_{10}(x)$ with $x$ being the non-diffractive gluon longitudinal momentum fraction (left panel) and the diffractive gluon longitudinal momentum fraction with respect to the proton (right panel) for single-diffractive production at $\sqrt{s} = 13$ TeV. Results for the LO collinear (black long-dashed) and the $k_{t}$-factorization (red solid) approaches are compared.
}
 \label{fig:logx_kTcoll}
\end{figure}

Figure ~\ref{fig:logx_kTcoll} shows the differential cross section as a function of $\log_{10}(x)$ where $x$ is defined as the longitudinal momentum fraction of proton carried by the gluon from non-diffractive side (left panel) or as the longitudinal momentum fraction of proton carried by the diffractive gluon emitted from pomeron/reggeon on diffractive side (right panel). In the case of non-diffractive gluon (left panel) we see that for extremely small $x$ values the LO collinear predictions strongly exceed the ones of the $k_{t}$-factorization. This effect also affects the rapidity spectra in the very forward/backward regions (see Fig.~\ref{fig:ypt_kTcoll}) and is partially related to a very poor theoretical control of the collinear PDFs in the range of $x$ below the vaule of $10^{-5}$.  

\begin{figure}[!htbp]
\begin{minipage}{0.47\textwidth}
 \centerline{\includegraphics[width=1.0\textwidth]{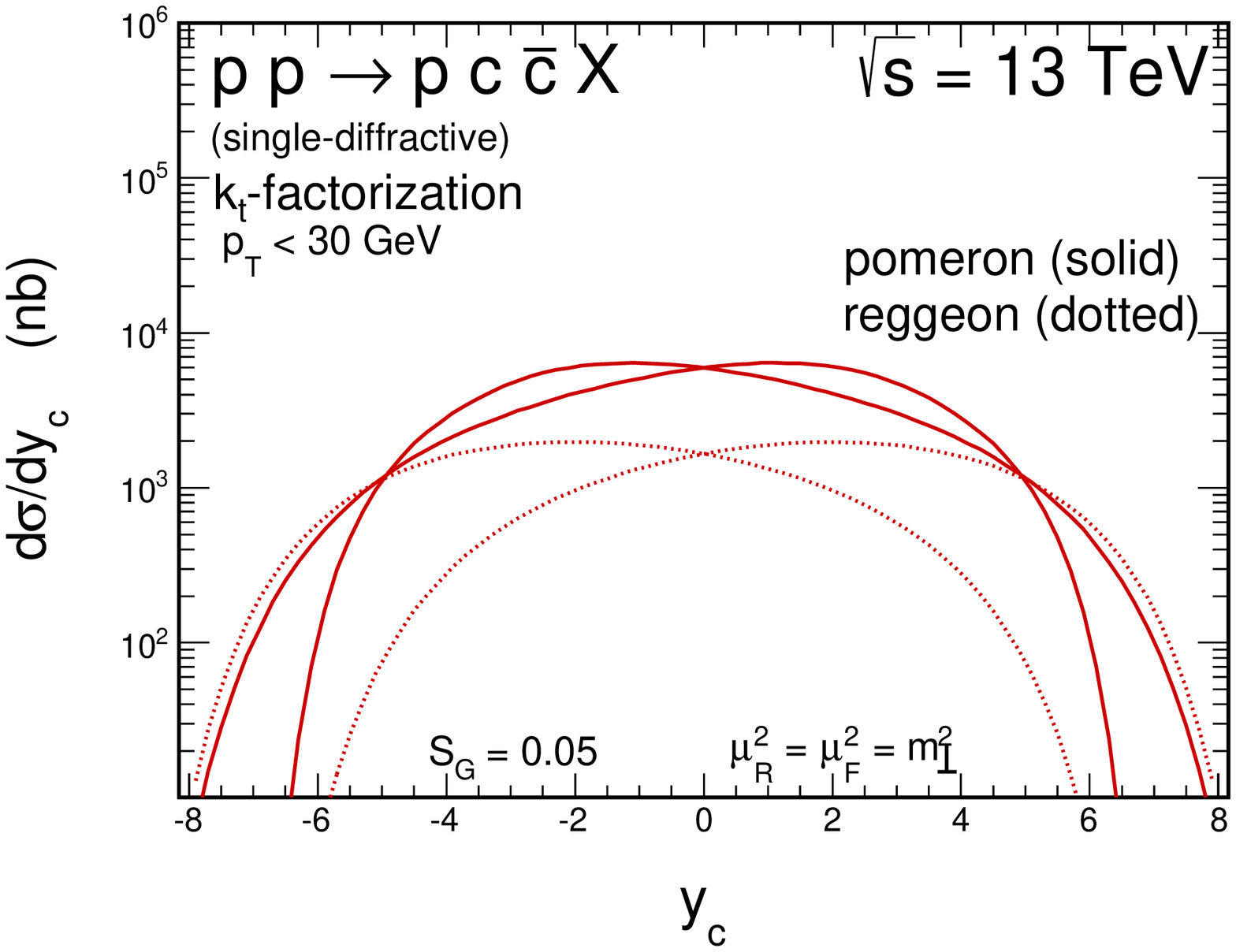}}
\end{minipage}
\hspace{0.5cm}
\begin{minipage}{0.47\textwidth}
 \centerline{\includegraphics[width=1.0\textwidth]{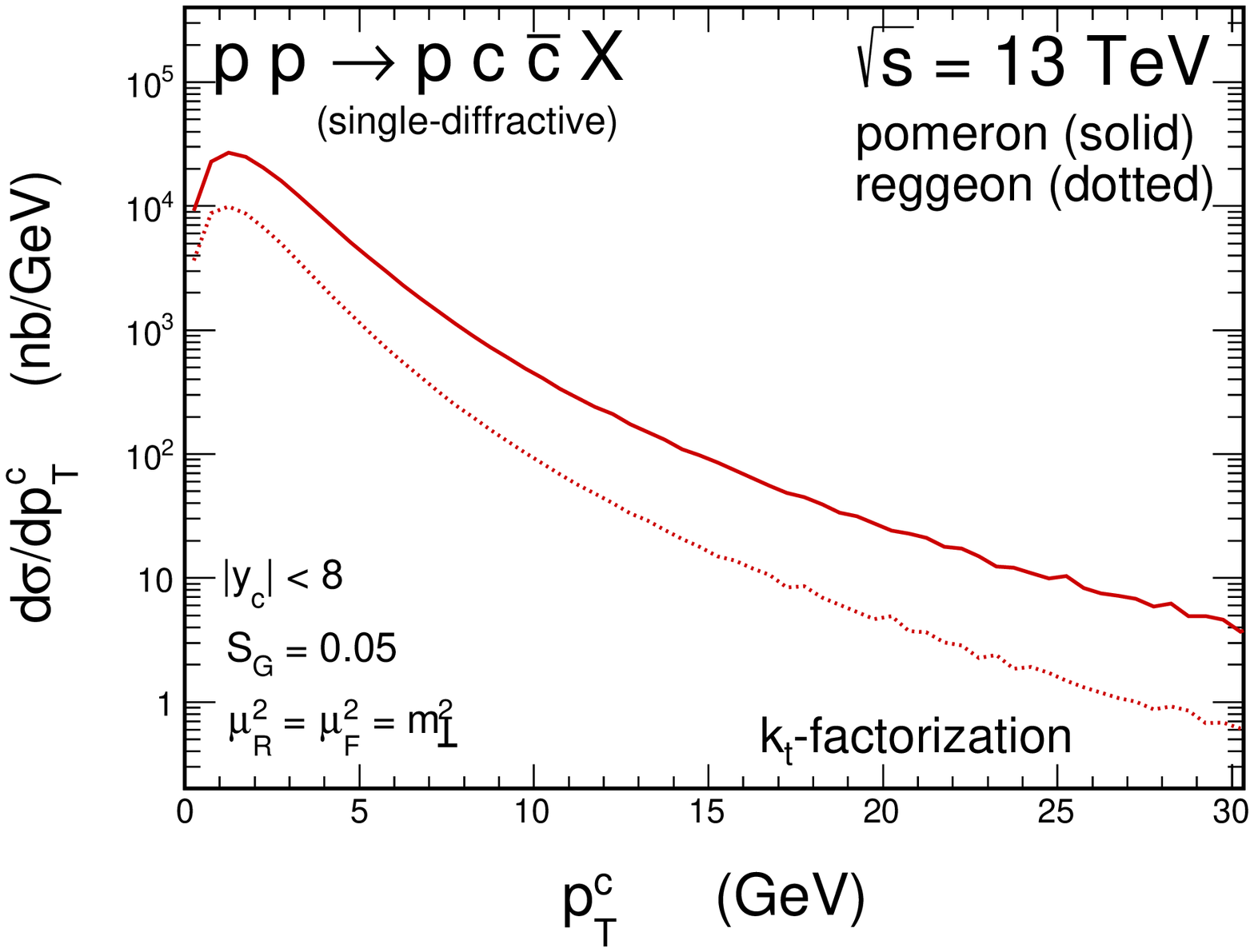}}
\end{minipage}
   \caption{
\small Rapidity (left panel) and transverse momentum (right panel) distributions of $c$ quarks (antiquarks) for single-diffractive production at $\sqrt{s} = 13$ TeV calculated with the $k_{t}$-factorization approach. Contributions of the $g(I\!P)\operatorname{-}g(p)$, $g(p)\operatorname{-}g(I\!P)$, $g(I\!R)\operatorname{-}g(p)$, $g(p)\operatorname{-}g(I\!R)$ mechanisms are shown separately.
}
 \label{fig:ypt_kT_PR}
\end{figure}

In Fig.~\ref{fig:ypt_kT_PR} we show again the rapidity (left panel) and transverse momentum (right panel) distributions of $c$ quarks (antiquarks) calculated in the $k_{t}$-factorization approach. Here contributions from the pomeron and the reggeon exchanges are shown separately. The estimated sub-leading reggeon contribution is of similar size as the one of the leading pomeron. In the single-diffractive case the maxima of rapidity distributions for $g(I\!P)\operatorname{-}g(p)$ and $g(p)\operatorname{-}g(I\!P)$ (or $g(I\!R)\operatorname{-}g(p)$ and $g(p)\operatorname{-}g(I\!R)$) mechanisms are shifted to forward and backward rapidities with respect to the non-diffractive case. This is related to the upper limit on diffractive gluon longitudinal momentum fraction ($x \leq x_{I\!P}$).  

Now we start our presentation of correlation observables. They cannot be calculated within the LO collinear factorization but are easily obtainable in the $k_{t}$-factorization approach.

\begin{figure}[!htbp]
\begin{minipage}{0.47\textwidth}
 \centerline{\includegraphics[width=1.0\textwidth]{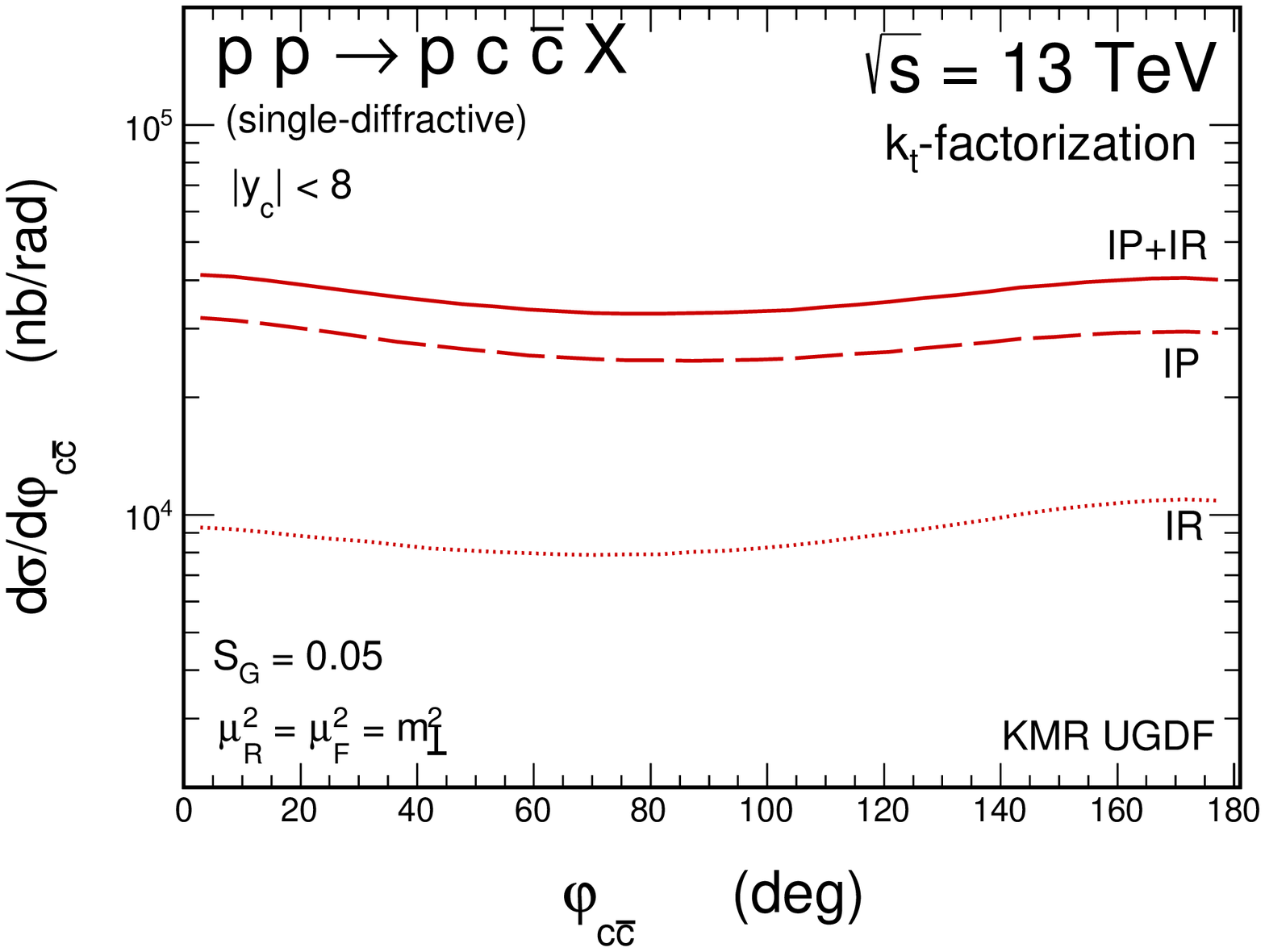}}
\end{minipage}
\hspace{0.5cm}
\begin{minipage}{0.47\textwidth}
 \centerline{\includegraphics[width=1.0\textwidth]{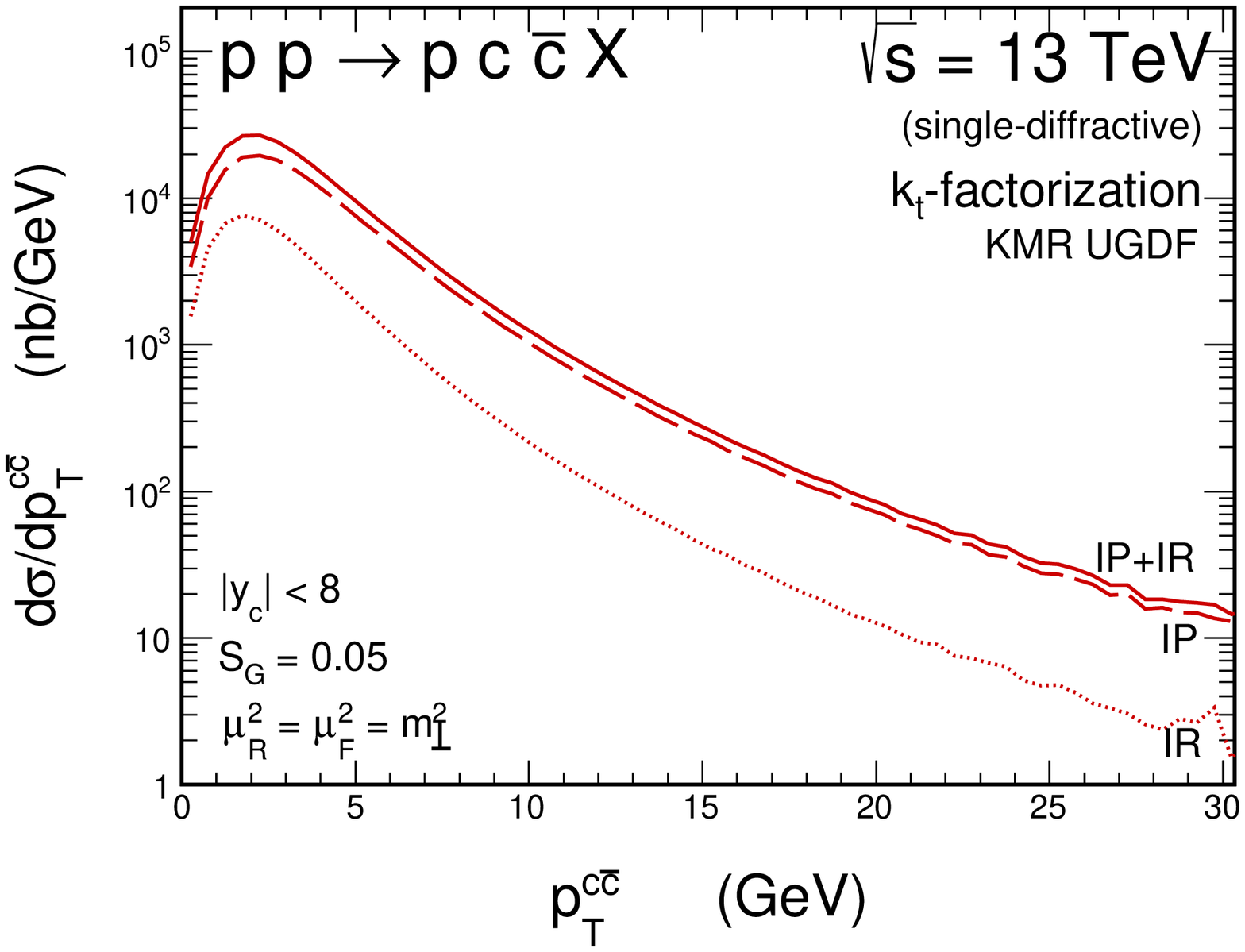}}
\end{minipage}
   \caption{
\small The distribution in $\phi_{c \bar c}$ (left panel) and distribution in $p_{T}^{c\bar c}$ 
(right panel) for $k_{t}$-factorization at $\sqrt{s}$ = 13 TeV.
}
 \label{fig:phid_ptsum_kT_PR}
\end{figure}

The distribution of azimuthal angle $\varphi_{c \bar c}$ between $c$ quarks and $\bar c$ antiquarks is shown in the left panel of Fig.~\ref{fig:phid_ptsum_kT_PR}. The $c \bar c$ pair transverse momentum distribution $p^{c \bar c}_{T} = |\vec{p^{c}_{t}} + \vec{p^{\overline{c}}_{t}}|$ is shown in the right panel. The results of the full phase-space calculations illustrate that the quarks and antiquarks in the $c \bar c$ pair are almost uncorrelated in the azimuthal angle between them and are often produced in the configuration with quite large pair transverse momenta. The distributions may be different if one includes the kinematical cuts related to the experimental coverage of detectors and/or hadronization effects. Exact calculation of the absorptive corrections may also have some influence on the shapes of the distributions (especially for $\varphi_{c \bar c}$). However, in the moment, including them is technically a difficult task and goes beyond the scope of this paper.      

\begin{figure}[!htbp]
\begin{minipage}{0.4\textwidth}
 \centerline{\includegraphics[width=1.0\textwidth]{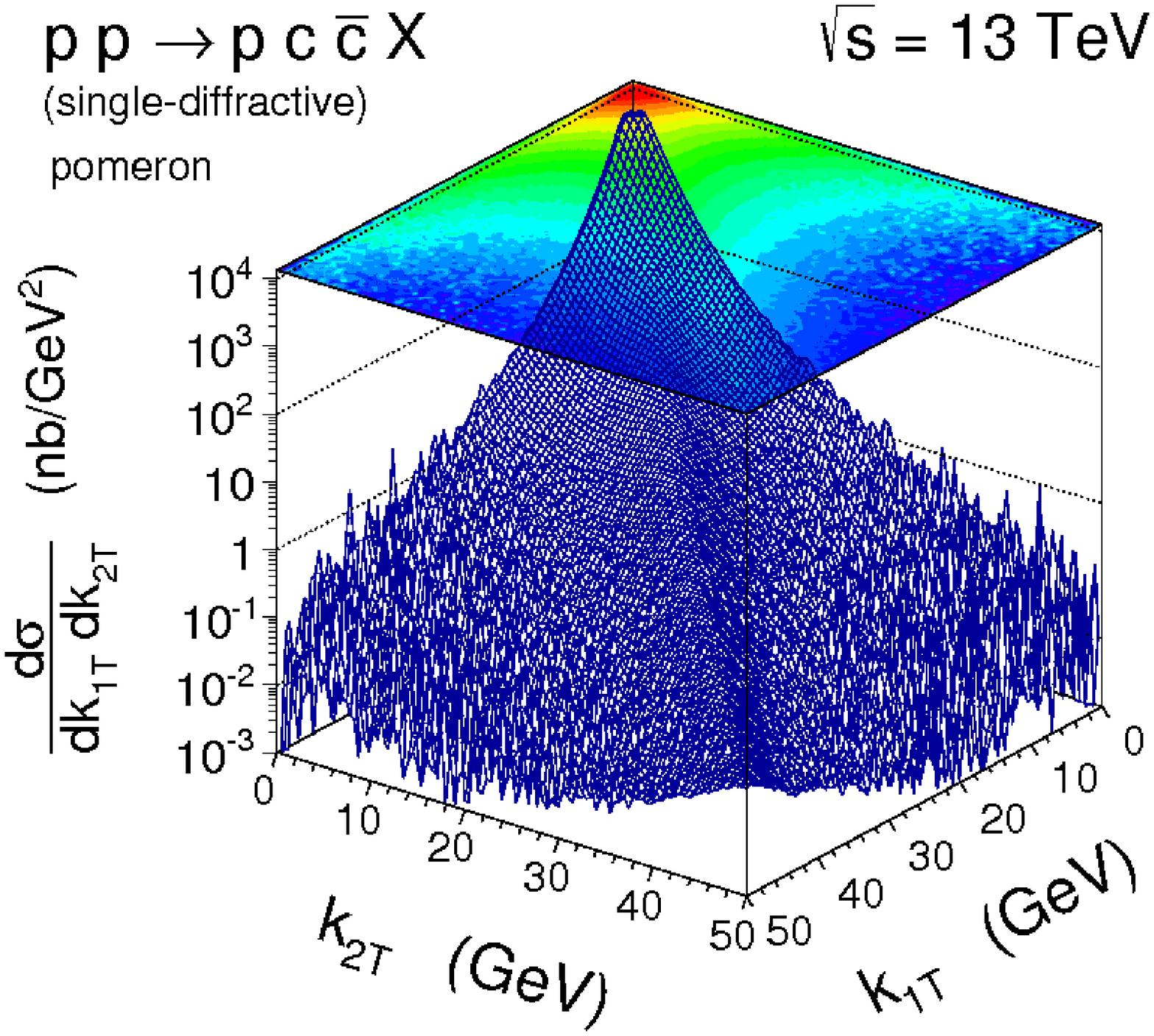}}
\end{minipage}
\hspace{0.5cm}
\begin{minipage}{0.4\textwidth}
 \centerline{\includegraphics[width=1.0\textwidth]{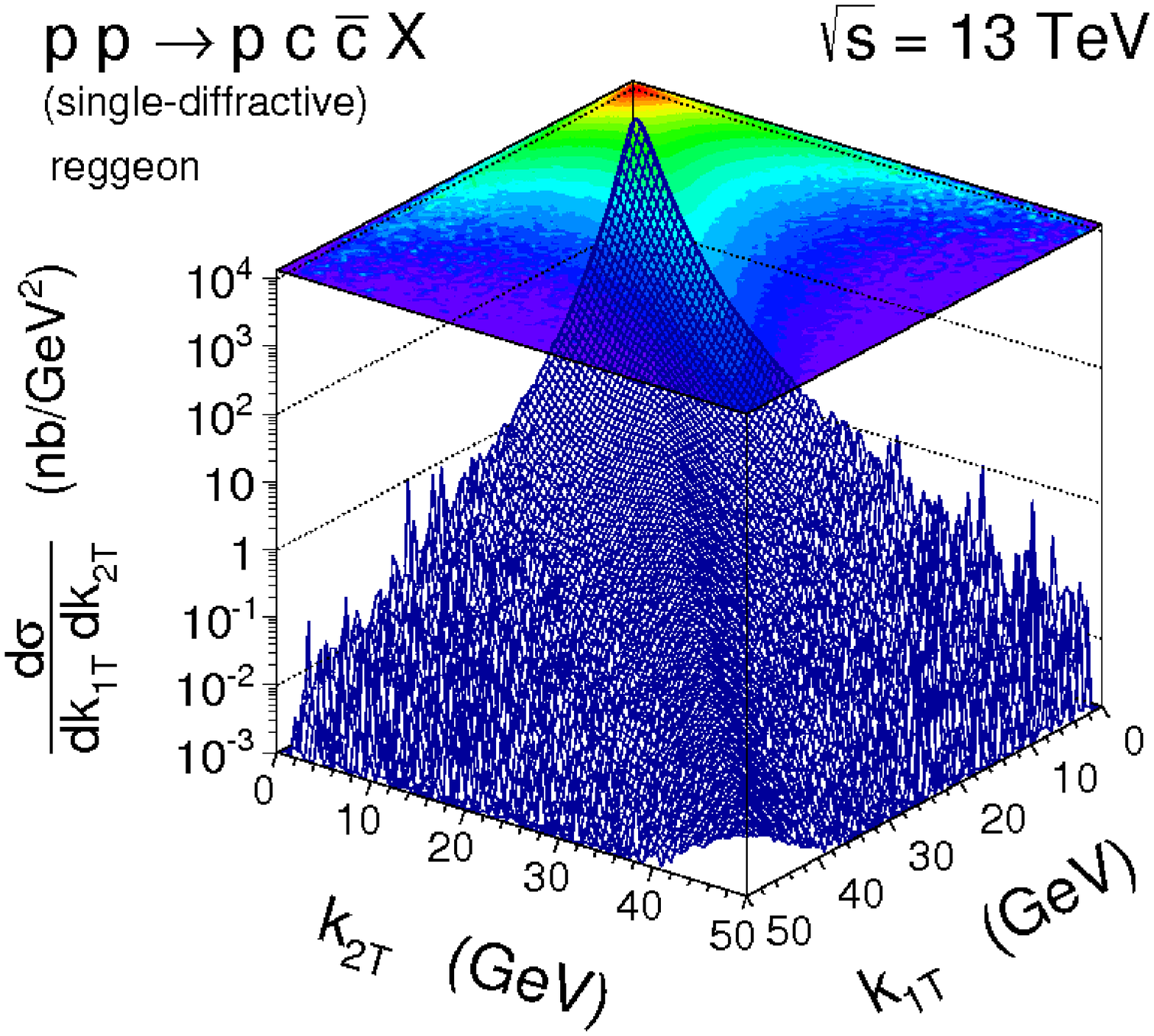}}
\end{minipage}
   \caption{
\small Double differential cross sections as a function of initial gluons transverse momenta $k_{1T}$ and $k_{2T}$ for single-diffractive production of charm at $\sqrt{s}=13$ TeV. The left and rights panels correspond to the pomeron and reggeon exchange mechanisms, respectively. 
}
 \label{fig:q1tq2t_kT_PR}
\end{figure}
\begin{figure}[!htbp]
\begin{minipage}{0.4\textwidth}
 \centerline{\includegraphics[width=1.0\textwidth]{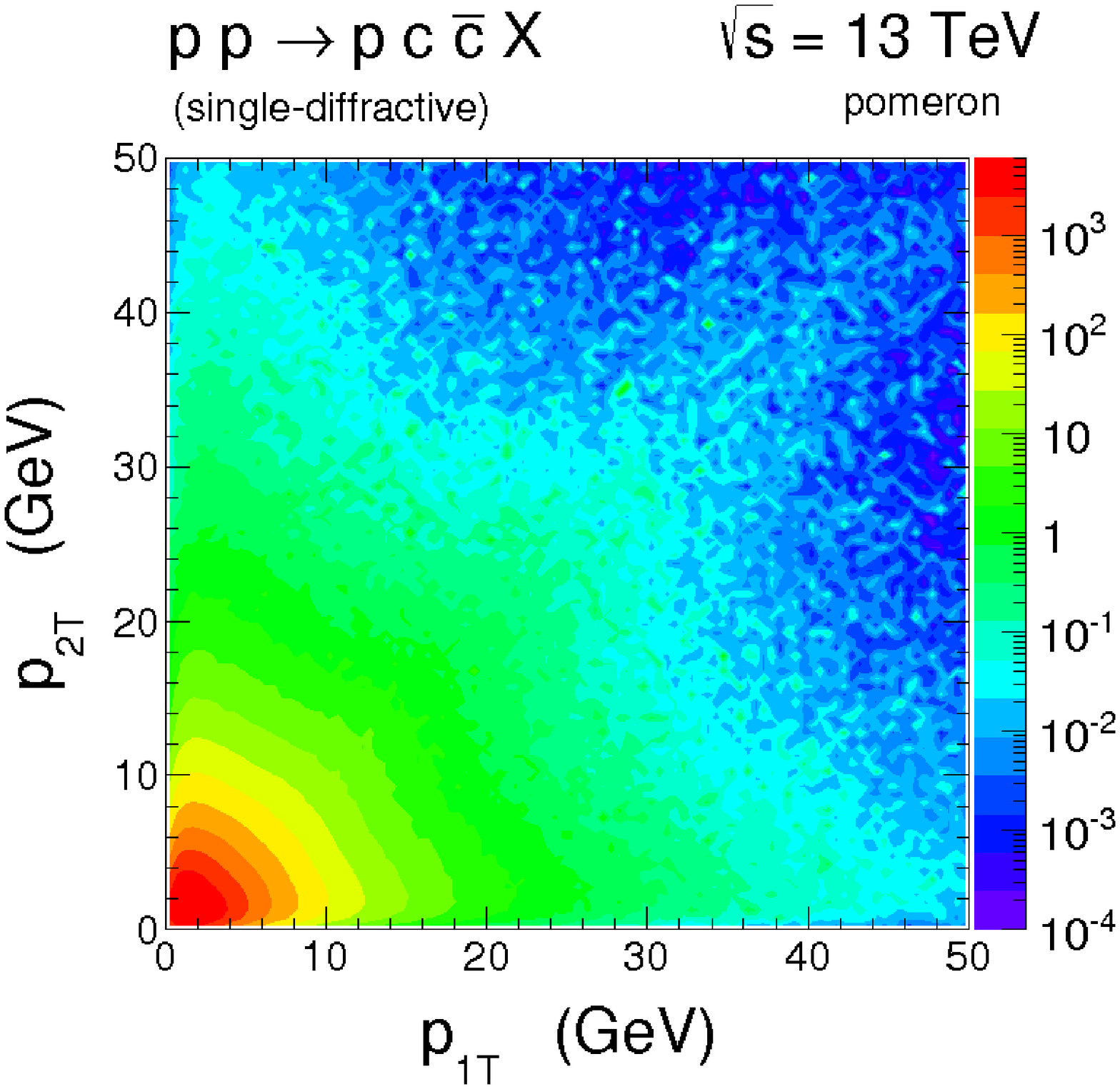}}
\end{minipage}
\hspace{0.5cm}
\begin{minipage}{0.4\textwidth}
 \centerline{\includegraphics[width=1.0\textwidth]{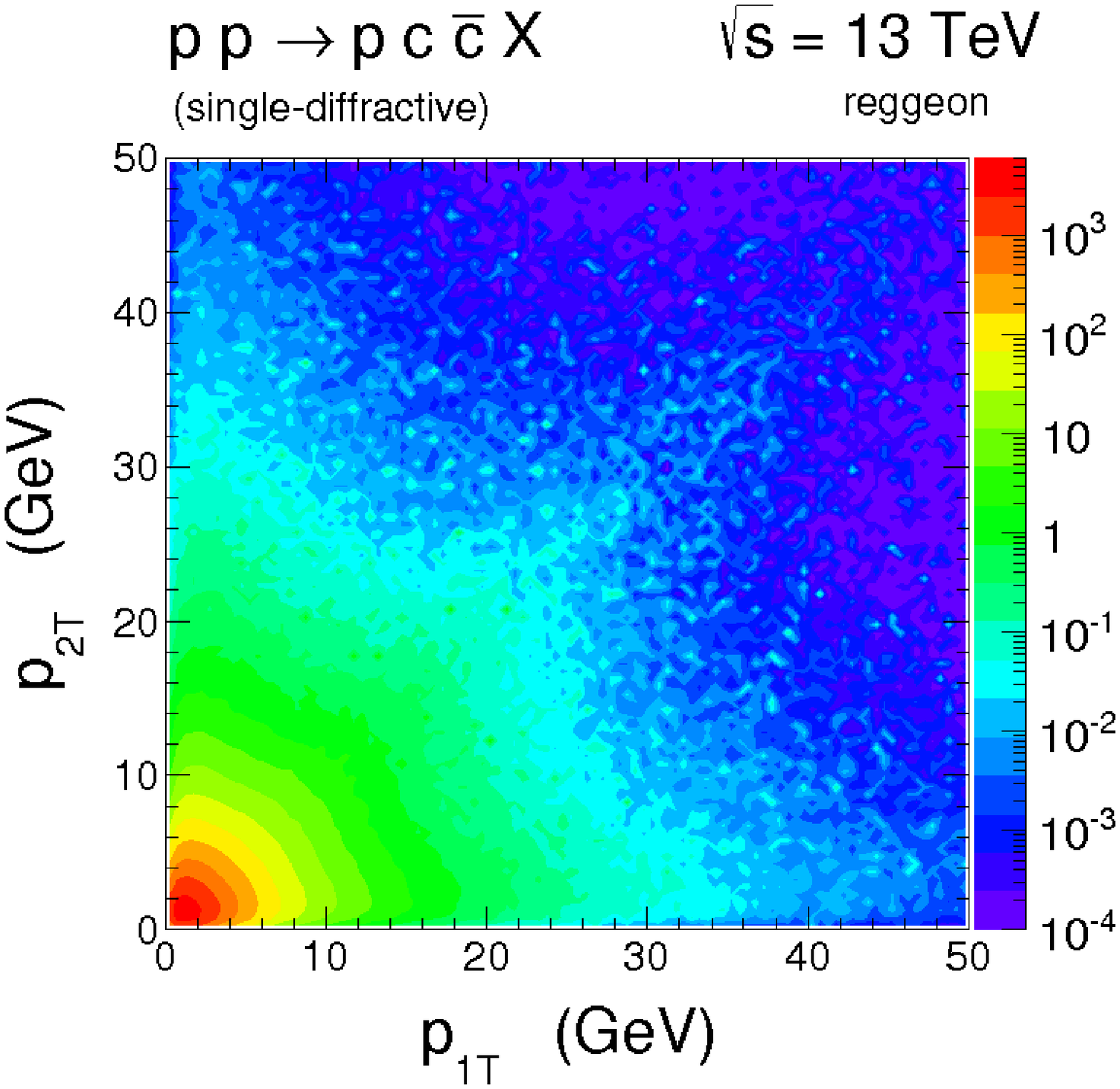}}
\end{minipage}
   \caption{
\small  Double differential cross sections as a function of transverse momenta of outgoing $c$ quark $p_{1T}$ and outgoing $\bar c$ antiquark $p_{2T}$ for single-diffractive production of charm at $\sqrt{s}=13$ TeV. The left and rights panels correspond to the pomeron and reggeon exchange mechanisms, respectively. 
}
 \label{fig:p1tp2t_kT_PR}
\end{figure}

Figures~\ref{fig:q1tq2t_kT_PR} and \ref{fig:p1tp2t_kT_PR} show the double differential cross sections as a functions of transverse momenta of incoming gluons ($k_{1T}$ and $k_{2T}$) and transverse momenta of outgoing $c$ and $\bar c$ quarks ($p_{1T}$ and $p_{2T}$), respectively. We observe that quite large incident gluon transverse momenta enter into our calculations. The major part of the cross section is concentrated in the region of small $k_{t}$'s of both gluons but long tails are present. Transverse momenta of the outgoing particles are not balanced as they are in the case of the LO collinear approximation. Such asymmetric configurations, where one $p_{t}$ is small and the second one is large, correspond to the higher-order corrections in the collinear case and are also present in the $k_{t}$-factorization approach. Both pomeron and reggeon components give similar correlations in these planes.

\begin{figure}[!htbp]
\begin{minipage}{0.4\textwidth}
 \centerline{\includegraphics[width=1.0\textwidth]{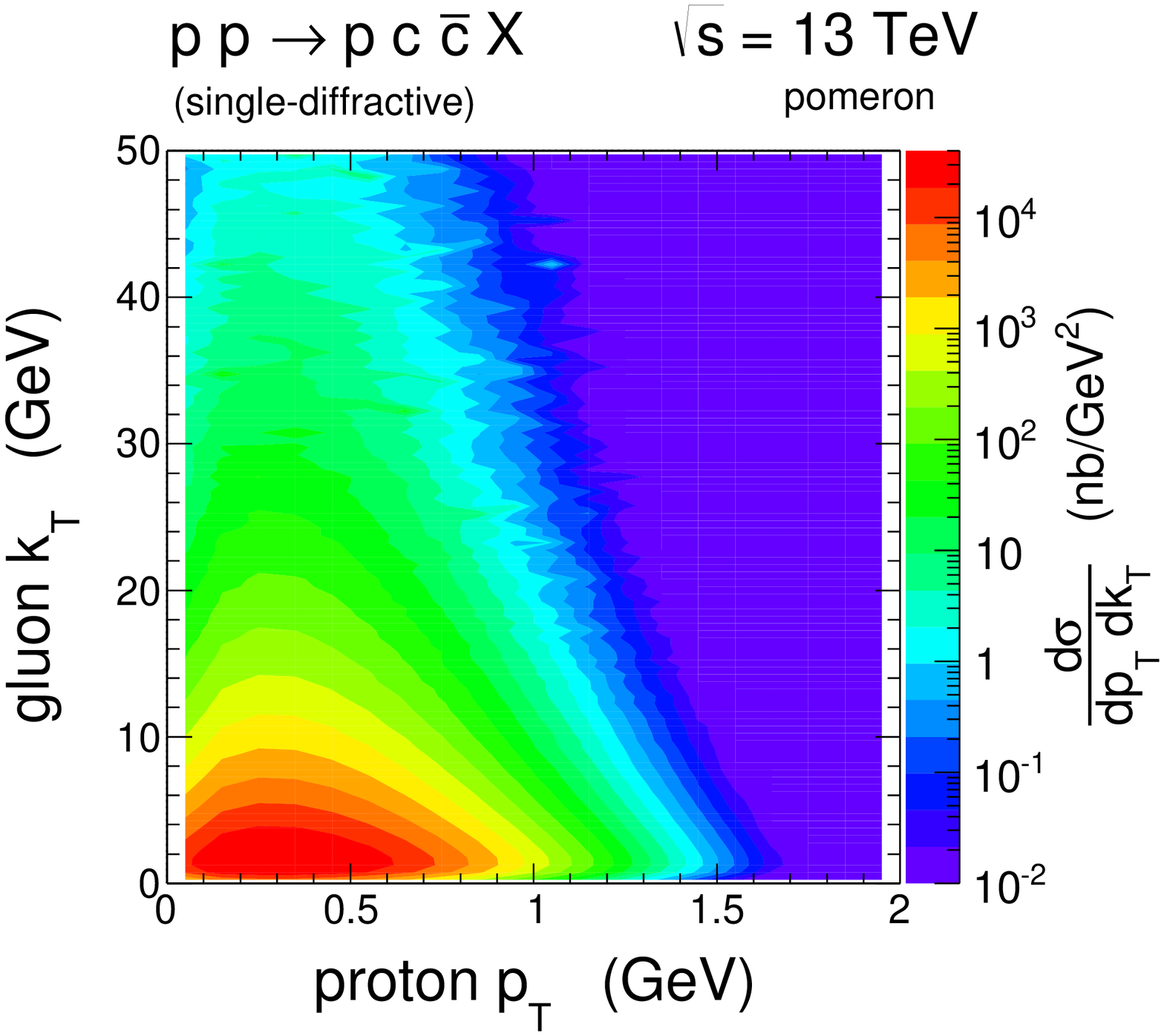}}
\end{minipage}
\hspace{0.5cm}
\begin{minipage}{0.4\textwidth}
 \centerline{\includegraphics[width=1.0\textwidth]{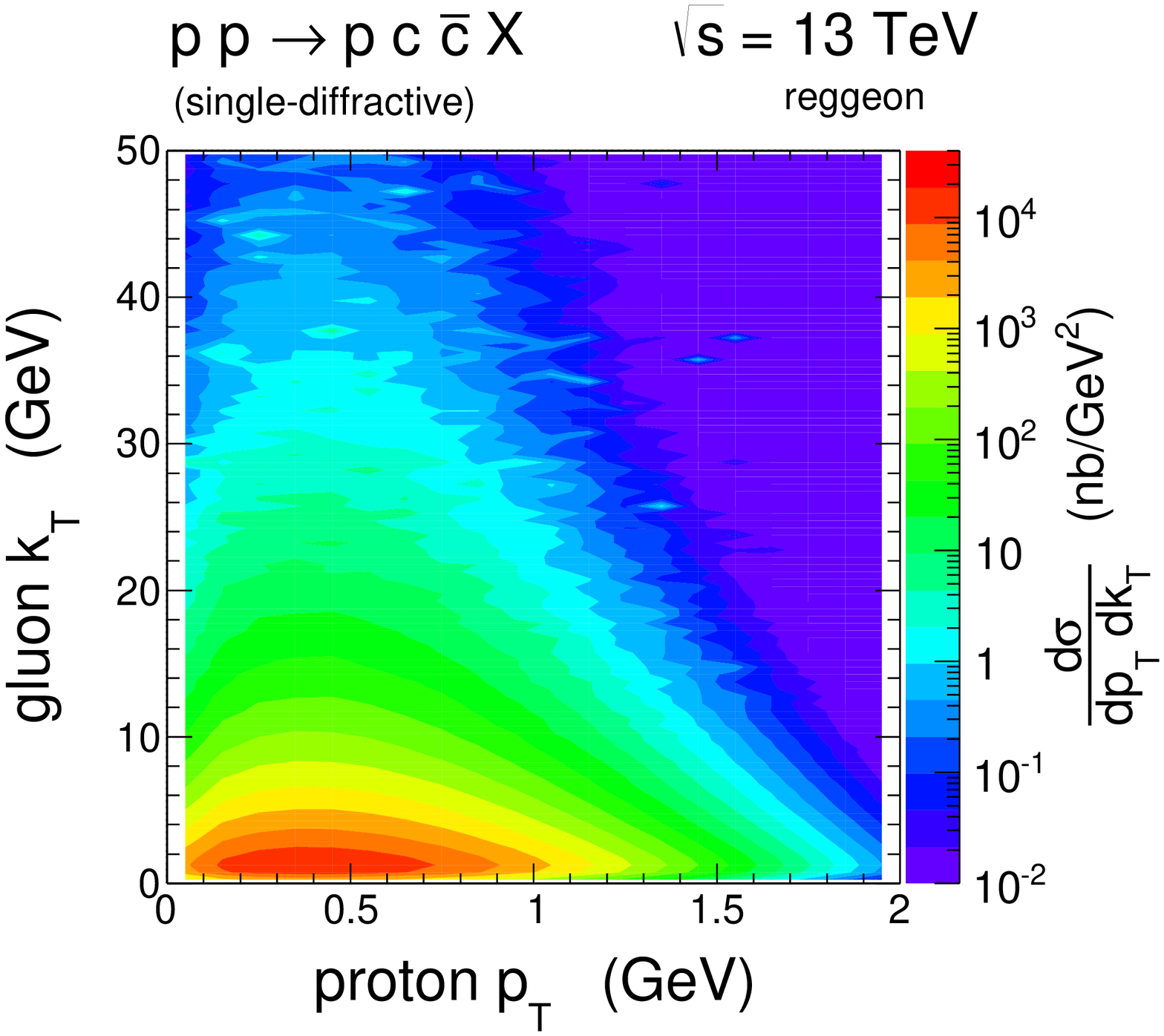}}
\end{minipage}
   \caption{
\small Double differential cross sections as a function of $p_{T}$ of outgoing proton and $k_{T}$ of incident gluon for the single-diffractive production of charm at $\sqrt{s}=13$ TeV. Left and rights panels correspond to the pomeron and reggeon exchange mechanisms, respectively. 
}
 \label{fig:qtptproton_kT_PR}
\end{figure}

In Fig.~\ref{fig:qtptproton_kT_PR} we show a double differential cross section as a function of the transverse momenta of outgoing proton and incident gluon (on the pomeron side). Again we see quite large gluon transverse momenta, significantly larger than the one of the outgoing proton. The cross section is concentrated in the region of proton $p_T$ smaller than 1~GeV. From the kinematics, the transverse momentum of the outgoing proton must be equal to the transverse momentum of the pomeron (or reggeon). This confirms that, in the first approximation, the pomeron (or reggeon) transverse momentum may be neglected and one can assume its zero-influence on the transverse momentum of the gluon emitted from the pomeron (or reggeon). For completeness this effect could be included in a future. We leave this for our future studies.   

\begin{figure}[!htbp]
\begin{minipage}{0.4\textwidth}
 \centerline{\includegraphics[width=1.0\textwidth]{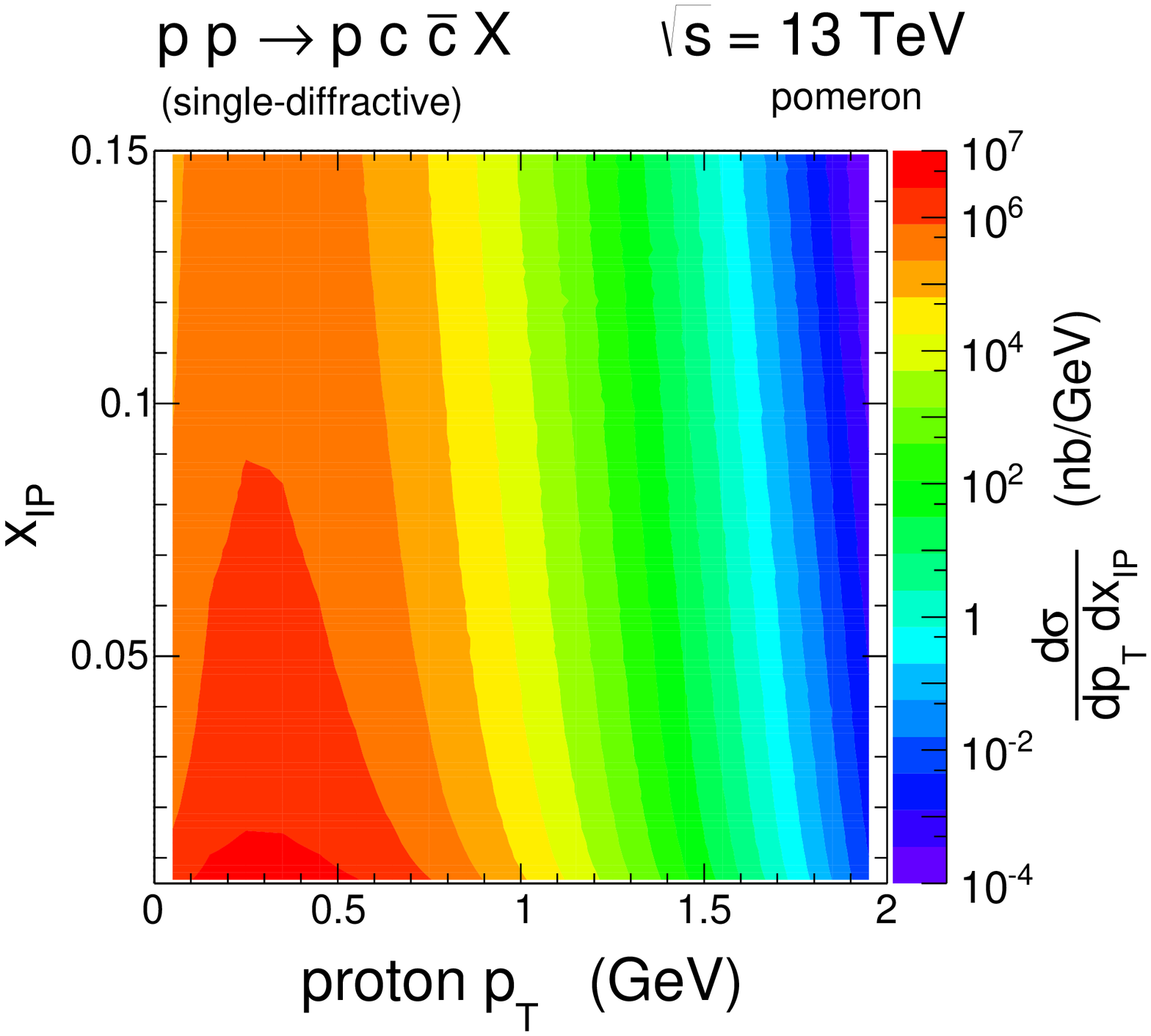}}
\end{minipage}
\hspace{0.5cm}
\begin{minipage}{0.4\textwidth}
 \centerline{\includegraphics[width=1.0\textwidth]{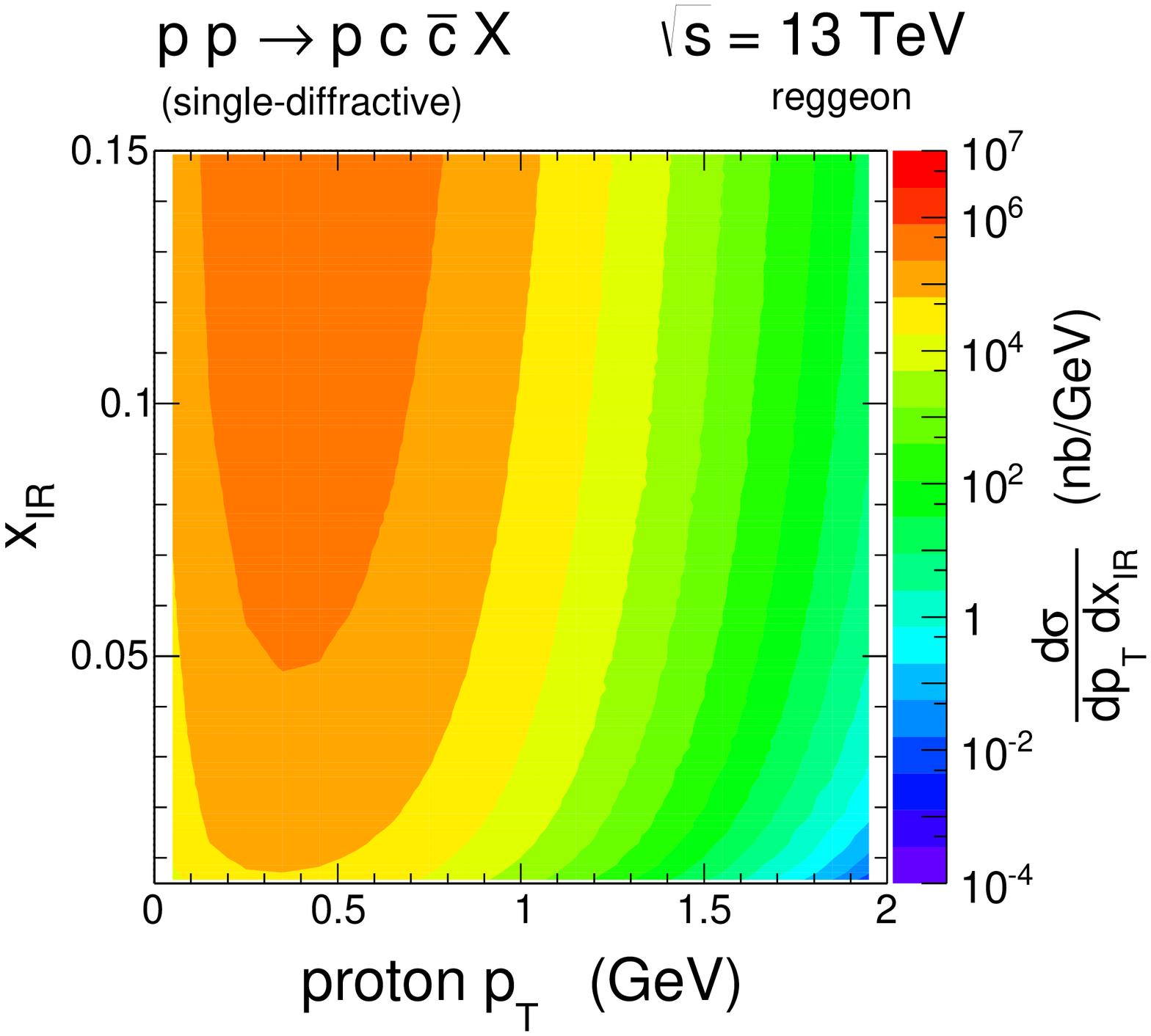}}
\end{minipage}
   \caption{
\small  Double differential cross sections as a function of $p_{T}$ of outgoing proton and longitudinal momentum fraction of the pomeron $x_{I\!P}$ (left panel) and of the reggeon $x_{I\!R}$ (right panel) for single-diffractive production of charm at $\sqrt{s}=13$ TeV.
}
 \label{fig:xpompt_kT_PR}
\end{figure}

Figure ~\ref{fig:xpompt_kT_PR} shows double differential cross sections as a function of $p_{T}$ of outgoing proton and longitudinal momentum fraction of pomeron $x_{I\!P}$ (left panel) and reggeon $x_{I\!R}$ (right panel). The maxima of the cross sections for pomeron and reggeon contributions are concentrated in different regions of longitudinal momentum fractions. The pomeron contribution is strongly peaked at very small $x_{I\!P}$'s while the reggeon component rises when going to larger $x_{I\!R}$'s. However, the observed differences are not enough to show a clear way for experimental distinction between the both mechanisms. Similar conclusions were obtained in the case of our former studies based on the collinear approach \cite{Luszczak:2014cxa}.

\begin{figure}[!htbp]
\begin{minipage}{0.47\textwidth}
 \centerline{\includegraphics[width=1.0\textwidth]{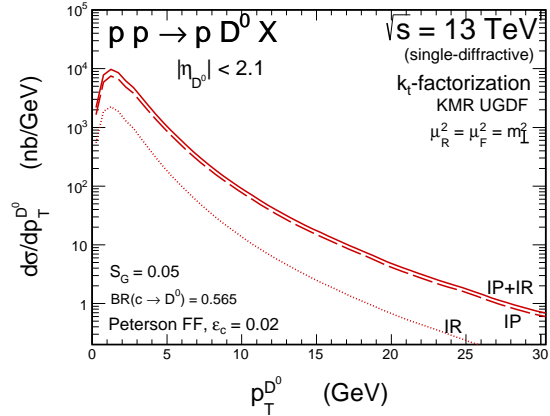}}
\end{minipage}
   \caption{
\small Transverse momentum distribution of $D^{0}$ meson within the ATLAS acceptance for single-diffractive production calculated with the $k_{t}$-factorization approach. Details are specified on the figure legend and in the text.
}
 \label{fig:ptD_ATLAS}
\end{figure}

Now we pass to the experimentally more motivated case of charm meson production. In Fig.~\ref{fig:ptD_ATLAS}, as an example, we present theoretical predictions for single-diffractive production of $D^{0}$ meson at the LHC. Here the $c \to D$ hadronization effects were taken into account with the help of the standard Peterson fragmentation function (for more details see \textit{e.g}.~\cite{Maciula:2013wg}). The fragmentation function is normalized with the experimentally well-known fragmentation fraction BR($c \to D^{0}) = 0.565$. From the different species of $D$ mesons the pseudoscalar neutral ones are produced most frequently. Here we concentrate on the ATLAS detector acceptance so the relevant cuts $p_{t}^{D} > 3.5$ GeV and $|\eta^{D}| < 2.1$ are applied. Both, pomeron and reggeon contributions are shown separately and the latter is found to be non-negligible, of the order of $\frac{I\!P}{I\!P + I\!R} \approx 20\%$. The relative reggeon contribution may be further slightly enhanced by increasing the lower limit of the $x_{I\!P}, x_{I\!R}$ but this will also result in a some (factor 2-3) reduction of the overall visible cross section. Within the full acceptance, the integrated cross section for the ATLAS detector is predicted at the level of $3-4$ $\mu$b which is quite large. However, for more definite conclusions whether the single-diffractive production of charm can be measured or not, a detailed feasibility studies are needed and are presented in the next section.    

\section{Feasibility Studies}

The predictions described in this paper can be verified experimentally using the LHC data. The characteristic signatures of a charm meson production in a single-diffractive mode are: large rapidity gap and scattered proton. The former one can be used by the LHCb experiment which has a good acceptance in the forward region. The latter, discussed in detail in the following section, can be used by ATLAS and CMS/TOTEM as these experiments are equipped with a special forward proton detectors.

Diffractive protons are usually scattered at very low angles, \textit{i.e.} they are produced into the LHC beamline, thus are not visible. In order to measure them dedicated detectors, located in the LHC beam pipe hundreds of meters away from the interaction point, are needed. At the LHC there are two sets of such detectors: TOTEM/CT-PPS~\cite{TOTEM} installed around CMS interaction point and ALFA~\cite{ALFA} and AFP~\cite{AFP} which are a part of the ATLAS experiment. Studies presented in this section focus on the ATLAS detectors, but the results are expected to be very similar for the CMS/TOTEM case.

Charmed mesons are identified using tracks reconstructed by the ATLAS inner detector. Diffractive signature is due to the forward proton tag. Taking this into account, there are two main sources of background: soft single-diffractive and non-diffractive production. The presence of forward proton in the latter case is usually due to the pile-up -- a situation, in which there are more than one interaction during a bunch-crossing.

\subsection{Charm Meson Reconstruction}
Studies performed below are similar to the ones done by ATLAS (see Ref. \cite{ATLAS_D_meson}). In the following, as an example, the production of $D^{*\pm}$ charmed mesons (hereafter called \textit{signal)} is discussed. However, it is worth stressing that the $D^{\pm}$ and $D^{\pm}_{s}$ productions are also feasible. Predictions are based on events generated accordingly to the theoretical calculations described earlier in this publication. In these studies hadronization of charm and anti-charm quarks was done using \textsc{Pythia 8} \cite{Pythia8}.

The response of the ATLAS detector was mimicked by applying a Gaussian smearing of $0.03 \otimes p_T$ in transverse momentum, $0.03$ in pseudorapidity and $0.02$ rad in azimuthal angle for each stable particle. These values are conservative and based on the performance studies published in \cite{ATLAS_JINST} and~\cite{ATLAS_reco_res}. In order to illustrate the detection efficiency, each track had a 85\% chance of being reconstructed (see \textit{e.g.} Ref.~\cite{ATLAS_reco_eff}). The reconstruction efficiency of forward proton detectors was set to 95\% \cite{AFP}.

The $D^{*\pm}$ mesons were identified in $D^{*\pm} \rightarrow D^0 \pi^{\pm} \rightarrow (K^-\pi^+)\pi^{\pm}$ decay channel. For each signal and background event, all pairs of oppositely-charged tracks, each with $p_T > 1.0$ GeV, were combined to form $D^0$ candidates. The decay length of the $D^0$ meson, calculated as the transverse distance between the candidate vertex and the primary vertex projected along the total transverse momentum of the track pair, was required to be greater than zero. The vertex reconstruction resolution of 300~$\mu$m was used \cite{ATLAS_reco_res}. In order to calculate the $D^0$ invariant mass, kaon and pion masses were assumed in turn for each track. In order to pass selection, events were required to be within $1.82 < M(D^0) < 1.91$~GeV mass window. Finally, $D^{*\pm}$ meson candidates were reconstructed in the range of transverse momentum $p_T^D > 3.5$ GeV and pseudorapidity $|\eta^D| < 2.1$. The background was reduced by requiring $p_T^D/E_T > 0.02$, where $E_T$ is a sum of the transverse energy of stable particles (except neutrinos) generated within the range of ATLAS calorimeter ($|\eta| < 4.9$).

\begin{figure}[!htbp]
  \centering
    \includegraphics[width=0.49\columnwidth]{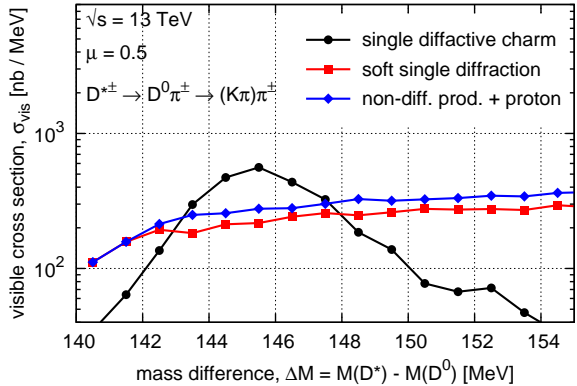}
\caption{Visible cross section after the event selection as a function of mass difference between $D^{*\pm}$ and $D^{0}$ mesons. Black line is for signal whereas red(blue) for single(non-) diffractive background. Pile-up intensity value was set to $\mu=0.5$.}
\label{fig_Dstar_signal}
\end{figure}

After the selection described above, a clear signal is visible (see Fig. \ref{fig_Dstar_signal}) in the region of $0.144 < \Delta M < 0.148$, where $\Delta M = M(K\pi\pi) - M(K\pi)$ is difference between the mass of $D^{*\pm}$ and $D^{0}$ meson. The figure was done for relatively small pile-up intensity value of $\mu = 0.5$.

\subsection{Forward Protons}
Due to the presence of the LHC magnets between the interaction point and forward proton detectors, the proton trajectory is not a straight line. Obviously, it depends on the settings of the magnetic field. In the simplest way, such settings, called \textit{optics}, could be characterized by the value of the betatron function at the interaction point, $\beta^*$. In the following, two optics at which ATLAS forward detectors could take data are considered: $\beta^* = 0.55$~m and $\beta^* = 90$~m. Their properties are described in details in Ref.~\cite{LHC_optics} and their choice is justified in Ref.~\cite{exclusive_jj_st}.

Scattered protons were assumed to be measured in the forward detectors: ALFA and AFP. Together with two considered optics, this means four different experimental conditions. Geometric acceptance for these cases is widely discussed in detail in Ref.~\cite{exclusive_jj_st}.

Protons were transported to the location of forward detectors using parametrised transport \cite{unfolding} trained on the relevant LHC optics \cite{LHC_optics}. Protons were checked to not be lost in the LHC aperture and to be within the detector active area. The beam-detector distance was set as in Ref.~\cite{exclusive_jj_st}.

\begin{figure}[!htbp]
  \includegraphics[width=0.49\columnwidth]{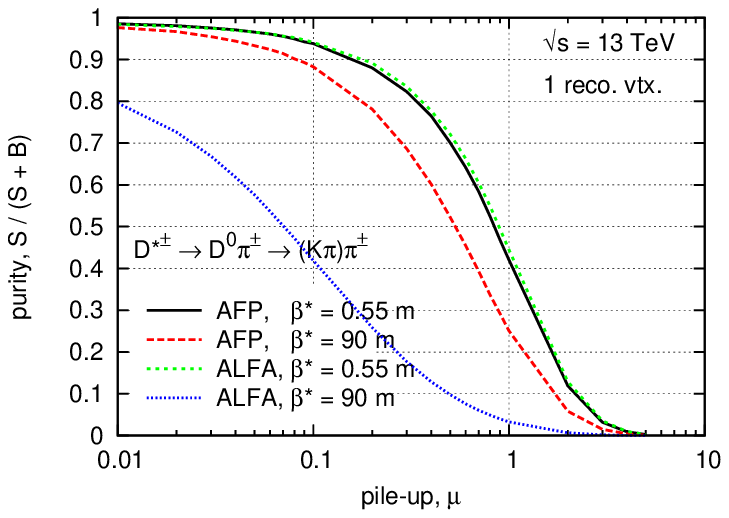}\hfill
  \includegraphics[width=0.49\columnwidth]{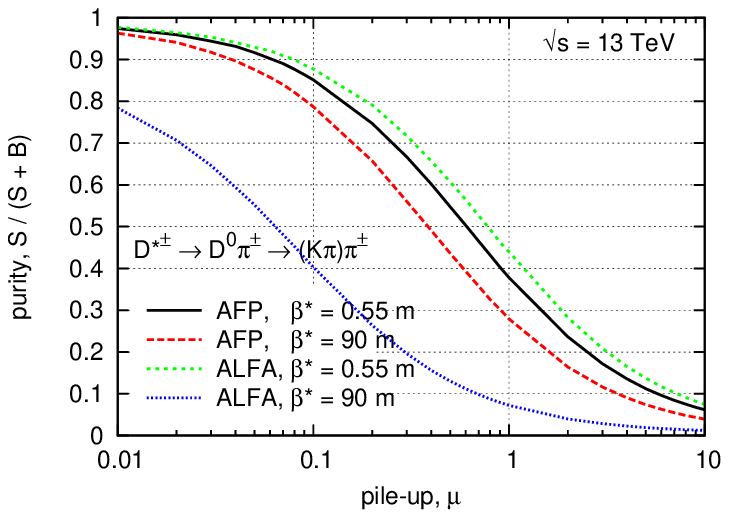}
\caption{Sample purity as a function of pile-up. The left panel shows situation with and the right panel without single vertex requirement (see text).}
\label{fig_Dstar_purity}
\end{figure}

The purity, defined as a ratio of signal $(S)$ to the sum of signal and background $(S + B)$ events, is shown in Fig. \ref{fig_Dstar_purity}. Left(right) figure is for the situation with(without) single vertex requirement. Smaller purity for the case of ALFA detector and $\beta^* = 90$~m optics is due to the protons scattered elastically. An additional anti-elastic selection (\textit{cf.} Ref.~\cite{ALFA_elastic_7TeV}) should improve the results.

In order to increase purity in the low pile-up data taking conditions, events with only one reconstructed vertex were considered. As is discussed in Ref.~\cite{exclusive_jj_st}, such selection will greatly reduce the combinatorial, non-diffractive background.

\subsection{Results}

\begin{figure}[!htbp]
  \includegraphics[width=0.49\columnwidth]{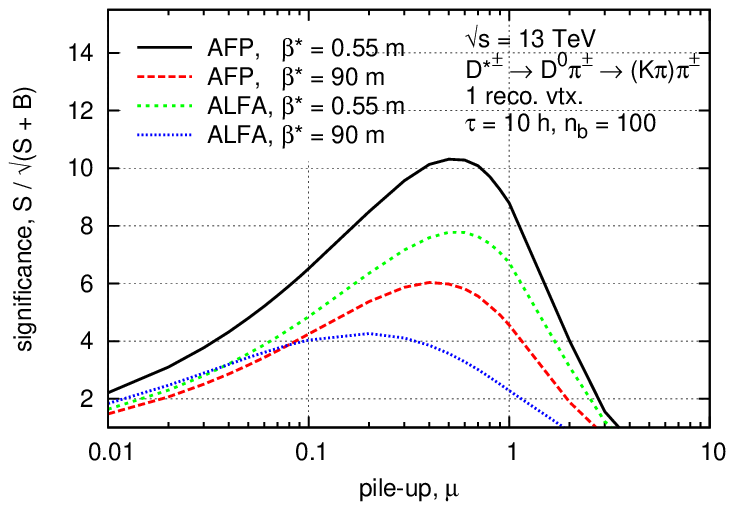}\hfill
  \includegraphics[width=0.49\columnwidth]{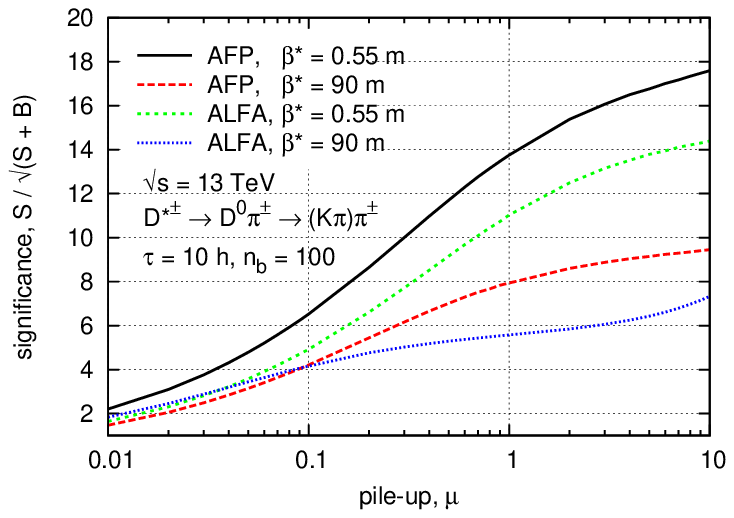}
\caption{Statistical significance for 10 hours of data-taking with 100 bunches as a function of pile-up intensity $\mu$. The left panel shows situation with and the right panel without single vertex requirement.}
\label{fig_Dstar_significance}
\end{figure}

The quality of the measurement can be expressed in terms of the statistical significance defined as $\frac{S}{\sqrt{S + B}}$. The results for all considered scenarios and the data-taking time of 10 hours with 100 colliding bunches as a function of pile-up is shown in Fig. \ref{fig_Dstar_significance}. Again, the left(right) figure is for the selection with(without) single vertex requirement.

As can be concluded from the above figures, pure and significant measurement can be performed for a wide range of experimental conditions ($0.05 < \mu < 1$). As expected, a single vertex requirement improves the purity for $\mu \lesssim 1$. For data taken with $\mu > 1$ the purity of the sample will be low, but the presence of signal should be evident.

\vspace{1cm}

\section{Conclusions}
Charm production is a good example where the higher-order QCD effects are very important. For the inclusive charm production we have shown that these effects can be effectively included in the $k_t$-factorization approach \cite{Maciula:2013wg}. In the present paper we have presented a first application of the $k_t$-factorization to the hard single-diffractive production.

In our approach we decided to use the so-called KMR method to calculate diffractive unintegrated gluon distribution. As usually in the KMR approach, we have calculated diffractive gluon UGDFs based on collinear distribution, which here is the diffractive collinear gluon distribution. In our calculations we have used the H1 Collaboration parametrization fitted to the HERA data on diffractive structure function and di-jet 
production.

Having obtained unintegrated diffractive gluon distributions we have performed calculations of several single-particle and correlation distributions. The results have been compared with the results of the leading-order collinear approximation. In general, the $k_t$-factorization approach leads to larger cross sections. However, the $K$-factor is strongly dependent on phase space point. We expect that our new predictions are better than the previous ones obtained in the collinear approach. Some correlation observables, like azimuthal angle correlation between $c$ and $\bar c$, and $c \bar c$ pair transverse momentum have been calculated for the first time.

The obtained cross sections for diffractive production of charmed mesons are fairly large and one could measure them. Such a measurement should give a chance not only to better understand the underlying mechanism of diffractive charm production but also the diffractive production in general. Therefore, we have supplemented our theoretical studies by a feasibility ones. The search of diffractive production of charm was assumed to be done using forward proton spectrometers installed by the ATLAS and TOTEM collaborations at the LHC. These experiments should be able to measure $D^{*\pm}$, $D^{\pm}$ and $D^{\pm}_{s}$ charmed mesons produced with diffractively scattered proton. Taking the example of $D^{*\pm}$ production, we have shown that after the signal selection, a pure and significant measurement is expected to be doable for a wide range of experimental conditions ($0.05 < \mu < 1$).

In future, other unintegrated diffractive parton distributions can be calculated in an analogous way (KMR method) as for gluons. Such a new distributions could be used for other hard diffractive processes. This would open a new situation in studying hard diffractive processes. Several useful correlation observables would become accessible with this technology. A next step would be \textit{e.g.} diffractive production of di-jets. For gauge boson production one could try to discuss transverse momentum distributions as was done for the inclusive case. Clearly, new perspectives are now open.   

\vspace{1cm}

{\bf Acknowledgements}

This study was partially supported by the Polish National Science Centre grants DEC-2013/09/D/ST2/03724 and DEC-2014/15/B/ST2/02528. Work of M.T. was partially supported by the Polish National Science Centre Mobility+ programme number 1285/MOB/IV/2015/0. 


\end{document}